\documentclass[review]{elsarticle}
 \pdfoutput=1
 
\usepackage{amsmath}
\usepackage{amsfonts}
\usepackage{algorithm}
\usepackage[noend]{algpseudocode}
\usepackage{graphicx}
\usepackage{float}
\usepackage{caption}
\usepackage{subcaption}
\usepackage{lineno,hyperref}
\modulolinenumbers[5]

\newcommand{\quotes}[1]{``#1''}
\newtheorem{theorem}{Theorem}

\begin{document}

\begin{frontmatter}

\title{Direct Computation of Two-Phase Icosahedral Equilibria of Lipid Bilayer Vesicles}

\author[add1]{Siming Zhao}
\ead{sz298@cornell.edu}
\author[add2]{Timothy Healey}
\ead{tjh10@cornell.edu}
\author[add3]{Qingdu Li}
\ead{liqd@cqupt.edu.cn}

\address[add1]{Sibley School of Mechanical and Aerospace Engineering, Cornell University, Ithaca, NY}
\address[add2]{Department of Mathematics, Cornell University, Ithaca, NY}
\address[add3]{Institue for Nonlinear Circuits and Systems, Chongqing University of Posts and Telecommunications, Chongqing, China}

\begin{abstract}
Correctly formulated continuum models for lipid-bilayer membranes present a significant challenge to computational mechanics. In particular, the mid-surface behavior is that of a 2-dimensional fluid, while the membrane resists bending much like an elastic shell. Here we consider a well-known \quotes{Helfrich-Cahn-Hilliard} model for two-phase lipid-bilayer vesicles, incorporating mid-surface fluidity, curvature elasticity and a phase field. We present a systematic approach to the direct computation of vesical configurations possessing icosahedral symmetry, which have been observed in experiment and whose mathematical existence has recently been established. We first introduce a radial-graph formulation to overcome the difficulties associated with fluidity within a conventional Lagrangian description. We use the so-called subdivision surface finite element method combined with an icosahedral-symmetric mesh. The resulting discrete equations are well-conditioned and inherit equivariance properties under a representation of the icosahedral group. We use group-theoretic methods to obtain a reduced problem that captures all icosahedral-symmetric solutions of the full problem. Finally we explore the behavior of our reduced model, varying numerous physical parameters present in the mathematical model.
\end{abstract}

\end{frontmatter}

\linenumbers

\section{Introduction}

Correctly formulated continuum models for lipid-bilayer membranes, exhibiting the properties of both fluids and solids, present a significant challenge to computational mechanics. At the molecular level, lipid molecules, each comprising a hydrophilic head and a hydrophobic tail, form double layers or bilayers under sufficient concentration. The heads coalesce on the two sides of the double-layer membrane, protecting the hydrophobic tails, which point inward toward the membrane mid-surface. The apparent freedom of the molecules to drift or exchange positions within the membrane is accounted for by mid-surface fluidity in the continuum model, while the mutual attraction of the heads on each of the lateral sides of the membrane leads to bending resistance – much like an elastic shell. The fluidity is captured elegantly via an Eulerian formulation, while the bending elasticity demands a Lagrangian description. However, the former is incomplete without knowledge of the current configuration, while the latter leads to grossly under-determined configurations.

In this work we consider a well-known “Helfrich-Cahn-Hilliard” model for two-phase lipid-bilayer vesicles, incorporating mid-surface fluidity, curvature elasticity and a phase field, cf. \cite{andelman1992equilibrium}, \cite{elliott2013computation}, \cite{jiang2000phase}, \cite{funkhouser2010dynamics}, \cite{ma2008viscous}, \cite{taniguchi1996shape}. In the absence of the latter, the model reduces to the well-known Helfrich model \cite{helfrich1973elastic}. The existence of a plethora of symmetry-breaking equilibria, bifurcating from the perfect spherical shape, has been recently established for this class of phase-field models \cite{healey2015symmetry}. The results include configurations possessing icosahedral symmetry, which have been observed in experiments sometimes taking on rather surprising \quotes{soccer-ball} shapes \cite{baumgart2003imaging}. Our aim here is to directly compute such configurations via symmetry methods and numerical bifurcation/continuation techniques \cite{healey1988group}. We point out that numerical gradient-flow techniques have been used to compute equilibria in two-phase models similar to that considered here \cite{elliott2013computation}, \cite{funkhouser2010dynamics}, \cite{ma2008viscous}, \cite{taniguchi1996shape}. This typically involves the addition of extra internal stiffness and damping mechanisms. Moreover, that approach constitutes a rather delicate and unsystematic procedure for obtaining specific equilibria, say, inspired by an experimentally observed configuration.  Certainly a great deal of patient, trial-and-error \quotes{tweaking} is required. Here we present a systematic approach to computing any equilibria within the multitude of symmetry types uncovered in \cite{healey2015symmetry}. We focus here on icosahedral symmetry, while methodically exploring parameter space via numerical continuation.

The outline of the work is as follows. In Section 2 we present the potential energy formulation of our problem, obtaining the weak form of the equilibrium equations in Lagrangian form. Due to the presence of curvature elasticity, an accurate finite-element model requires a $C^1$ formulation. As such, we employ the so-called subdivision surface finite element method, cf. \cite{cirak2000subdivision}, \cite{cirak2001fully}, which was first introduced for computer-graphics applications \cite{catmull1978recursively}. As pointed out in \cite{feng2006finite}, the resulting discrete equations are wildly ill-conditioned - a direct consequence of mid-plane fluidity. We get around this difficulty via the approach used in \cite{funkhouser2010dynamics}, \cite{healey2015symmetry}, \cite{treibergs1983embedded}, introducing the deformation as a radial graph over the unit sphere. This effectively eliminates the grossly under-determined mid-plane deformation, leading to a well-conditioned discretized system.

Presuming a mesh with icosahedral symmetry, we present the symmetry-reduction arguments in Section 4: The energy is invariant under a group action, implying that the discrete equilibrium equations are equivariant. We then deduce a symmetry-reduced problem, implemented via a projection operator coming from group representation theory. The reduced problem captures all solutions of the full problem having icosahedral symmetry. In Section 5 we present our numerical results for the reduced problem. We obtain a veritable catalog of two-phase equilibria for various values of the parameters - all having icosahedral symmetry and all obtained via numerical continuation. Among these are several \quotes{soccer-ball} configurations.

\section{Formulation}
\noindent We begin with the following phase-field elastic-shell \textit{potential energy} for a vesicle, defined on the current configuration, denoted by $\Sigma$, presumed isomorphic to the unit sphere $S^2$:
\begin{equation}\label{energy}
E=\int_{\Sigma}\bigg(BH^{2}+\sigma\big(\frac{\epsilon}{2}\parallel\nabla_{\Sigma}\phi\parallel^2+W(\phi)\big)\bigg)ds-pV_\Sigma,
\end{equation}
subject to the constraints,
\begin{subequations}
\begin{equation}
\int_{\Sigma}ds=4\pi,\label{constraint1}
\end{equation}
\begin{equation}
\int_{\Sigma}(\phi-\mu)ds=0. \label{constraint2}
\end{equation}
\end{subequations}
The scalar field $\phi: \Sigma\to R$ represents the phase concentration field governing the phase transition, $\nabla_{\Sigma}\phi$ is the surface gradient on the current surface configuration $\Sigma$, $H$ denotes the mean curvature of the surface $\Sigma$, $B>0$ is the constant \textit{bending moduli}, $\epsilon>0$ is a small \quotes{interfacial parameter}, $p\ge0$ is the prescribed internal pressure, $V$ is the volume of vesicle enclosed by the surface $\Sigma$, $\mu>0$ represents the average phase concentration on the surface, $W(\cdot)$ is a double-well potential, and $\sigma$ is a control parameter balancing the curvature and phase contributions to the total energy. Our double-well potential is defined as
\begin{equation}
W(\phi)=(\phi-1)^2(\phi+1)^2.
\end{equation}

Let $\textbf{x}=\textbf{x}(s^1,s^2)$ denote a parametrization of $\Sigma$ in terms of curvilinear coordinates $(s^1,s^2)\in\Omega$. The covariant tangent vectors are given by
\begin{gather}
\textbf{a}_\alpha=\frac{\partial \textbf{x}}{\partial s^\alpha}:=\textbf{x}_{,\alpha},\quad \alpha=1,2,
\end{gather}
where Greek indices ranging from $1$ to $2$, with repeated indices implying summation. The reciprocal tangent vectors, denoted $\textbf{a}^\alpha$, satisfying $\textbf{a}^\alpha\cdot\textbf{a}_\beta=\delta^\alpha_\beta$, are determined by
\begin{equation}
\textbf{a}^\alpha=a^{\alpha\beta}\textbf{a}_\beta,
\end{equation}
with the covariant and contravariant components of the metric tensor defined as $a_{\alpha\beta}=\textbf{a}_\alpha\cdot\textbf{a}_\beta=a_{\beta\alpha}$ and $a^{\alpha\beta}=\frac{(-1)^{\alpha-\beta}}{a}a_{(3-\alpha)(3-\beta)}$, respectively.
\noindent The unit normal to the surface is given by
\begin{equation}
\textbf{n}=\textbf{a}_3=\frac{\textbf{a}_1\times\textbf{a}_2}{\sqrt{a}},
\end{equation}
where $a=det[a_{\alpha\beta}]=\parallel\textbf{a}_1\times\textbf{a}_2\parallel^2$.
\noindent The mean curvature can be expressed as
\begin{equation}
H=-\frac{1}{2}\textbf{a}^\alpha\cdot\textbf{n}_{,\alpha},\quad\textrm{with $\textbf{n}_{,\alpha}=-(\textbf{n}\cdot\textbf{a}_{\beta,\alpha})\textbf{a}^\beta$}.
\end{equation}

Introducing two Lagrange multipliers $\lambda_s$ and $\lambda_\phi$, associated with (\ref{constraint1}) and (\ref{constraint2}) respectively, the modified energy functional takes the form
\begin{multline}\label{total_energy}
\Pi=\int_{\Sigma}\bigg(BH^{2}+\sigma\big(\frac{\epsilon}{2}\parallel\nabla_{\Sigma}\phi\parallel^2+W(\phi)\big)\bigg)ds+\lambda_s(\int_{\Sigma}ds-4\pi)\\
+\lambda_\phi(\int_{\Sigma}\phi ds-4\pi\mu)-\frac{1}{3}p\int_{\Sigma}\textbf{x}\cdot\textbf{n}ds,
\end{multline}
and using the parametrization of $\Sigma$ yields
\begin{multline}\label{energy_ref}
\Pi=\int_\Omega\bigg(BH^{2}+\sigma\big(\epsilon e_g+W(\phi)\big)\bigg)\sqrt{a}ds+\lambda_s(\int_\Omega\sqrt{a}ds-4\pi)\\
+\lambda_\phi(\int_\Omega\phi\sqrt{a}ds-4\pi\mu)-\frac{1}{3}p\int_\Omega\textbf{x}\cdot\textbf{n}\sqrt{a}ds,
\end{multline}
where $\Omega$ is the coordinate domain, $e_g:=\frac{1}{2}\parallel\nabla_{\Sigma}\phi\parallel^2=a^{\alpha\beta}\phi_{,\alpha}\phi_{,\beta}=\frac{a_{22}}{2a}\phi_{,1}^2+\frac{a_{11}}{2a}\phi_{,2}^2-\frac{a_{12}}{a}\phi_{,1}\phi_{,2}$. We now take the first variation, leading to
\begin{multline}\label{rawFirstVariation}
\delta\Pi=\int_\Omega\bigg(2BH\delta H+\sigma\epsilon\delta e_g+\sigma W^{'}(\phi)\delta\phi-\frac{1}{3}p(\textbf{n}\cdot\delta\textbf{x}+\textbf{x}\cdot\delta\textbf{n})\bigg)\sqrt{a}ds\\
+\int_\Omega\big(BH^2+\sigma\big(\epsilon e_g+W(\phi)\big)-\frac{1}{3}p\textbf{x}\cdot\textbf{n}\big)\delta\sqrt{a}ds\\
+\lambda_s\int_\Omega\delta\sqrt{a}ds+\lambda_\phi\int_\Omega(\sqrt{a}\delta\phi+\phi\delta\sqrt{a})ds.
\end{multline}
The variation quantities above have the following forms:
\begin{align*}
&\delta\sqrt{a}=\sqrt{a}\textbf{a}^\alpha\cdot\delta\textbf{a}_\alpha,\\
&\delta H=-\frac{1}{2}\textbf{a}^\alpha\cdot(\delta\textbf{n})_{,\alpha}+\frac{1}{2}a^{\alpha\beta}\textbf{n}_{,\alpha}\cdot\delta\textbf{
a}_\beta,\\
&\delta\textbf{n}=-(\textbf{n}\cdot\delta\textbf{a}_\alpha)\textbf{a}^\alpha,
\end{align*}
\begin{multline*}
\delta e_g=-\frac{2e_g}{\sqrt{a}}\delta a+\frac{1}{2a}(\phi_{,1}^2\delta a_{22}+\phi_{,2}^2\delta a_{11}-2\phi_{,1}\phi_{,2}\delta a_{12})\\
+\frac{1}{a}\big(a_{22}\phi_{,1}(\delta\phi)_{,1}+a_{11}\phi_{,2}(\delta\phi)_{,2}-a_{12}(\phi_{,1}(\delta\phi)_{,2}+\phi_{,2}(\delta\phi)_{,1})\big),
\end{multline*}
where $\delta a=2a\textbf{a}^\alpha\cdot(\delta\textbf{x})_{,\alpha}$ and $\delta a_{\alpha\beta}=\textbf{a}_\beta\cdot(\delta\textbf{x})_{,\alpha}+\textbf{a}_\alpha\cdot(\delta\textbf{x})_{,\beta}$. After substituting the above variation quantities into (\ref{rawFirstVariation}) and rearranging, the first variation condition for (\ref{total_energy}) reads
\begin{equation}\label{first_variation}
\delta\Pi=\int_\Omega\big[\textbf{d}_\alpha\cdot(\delta\textbf{x})_{,\alpha}+\textbf{m}\cdot\delta\textbf{n}_{,\alpha}-\frac{p}{3}\textbf{n}\cdot\delta\textbf{x}+\epsilon\sigma\delta e_g+\big(\sigma W^{'}(\phi)+\lambda_\phi\big)\delta\phi\big]\sqrt{a}ds=0,
\end{equation}
where the following identities can be shown to hold:
\begin{align*}
&\textbf{d}_\alpha=BH(a^{\alpha\beta}\textbf{n}_{,\beta}+H\textbf{a}^\alpha)+\big(\lambda_s+\lambda_\phi\phi+\sigma W(\phi)+\epsilon\sigma e_g\big)\textbf{a}^\alpha-\frac{p}{3}\big((\textbf{x}\cdot\textbf{n})\textbf{a}^\alpha-(\textbf{x}\cdot\textbf{a}^\alpha)\textbf{n}\big),\\
&\textbf{m}_\alpha=-BH\textbf{a}^\alpha,\\
&\delta\textbf{n}_{,\alpha}=-\big(\textbf{n}\cdot(\delta\textbf{x})_{,\alpha}\big)\textbf{a}^\alpha.
\end{align*}

\section{Subdivision Surface Finite Element Method and Radial Graph Description}
\noindent Since the bending energy is a quadratic functional of the mean curvature, we require the approximation of the surface $\Sigma$ to be $C^1$. As such, we employ a recently developed $C^1$ thin-shell finite element procedure \cite{cirak2000subdivision,cirak2001fully}, based on the method of subdivision surface. The method was originally devised for rendering smooth surfaces in computer graphics \cite{catmull1978recursively}. The shape function for a node of the triangular mesh has support not only over the triangles connected to the node, also to adjacent triangles. Instead of \textit{interpolating} the surface, the method approximates a \textit{limit surface} which does not pass through the nodal points.

\noindent We adopt a Ritz-style finite element procedure similar to \cite{feng2006finite}. For $jth$ triangular element in the control mesh, we chose a local parametrization $(s^1,s^2)$ as two of its barycentric coordinates within their range
\begin{equation}\label{T}
\Omega_j\equiv\{(s^1,s^2),\quad\textrm{s.t. $s^\alpha\in[0,1]$,\quad$s^1+s^2\leq1$}\}.
\end{equation}
The triangle $\Omega_j$ in the $(s^1,s^2)$ plane can be regarded as a master element domain.
\begin{figure}[!htb]
    \centering
    \includegraphics[width=0.75\textwidth]{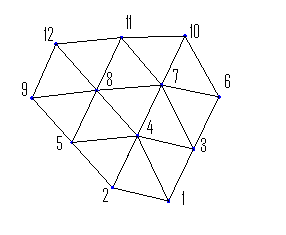}
    \caption{A regular box-spline patch with $12$ control points}
    \label{ringPatch}
\end{figure}

If the valencies of the nodes (number of connecting edges) of a given triangle are all equal to $6$, the resulting piece of limit surface is exactly described by a single box-spline patch, called a \textit{regular patch},  in Figure \ref{ringPatch}. If one of its nodes has valence other than $6$, the resulting patch is called \textit{irregular patch}. In order to do evaluation on irregular patches, further subdivisions are needed until evaluation points fall into subdivided regular patches. For details, refer to \cite{cirak2000subdivision}.

Figure \ref{ringPatch} shows a local numbering of the nodes lying in a generic element's (triangle $4-7-8$) nearest neighbourhood. The Loop's subdivision scheme \cite{loop1987smooth} leads to classical quartic box-splines. Therefore, the local parametrization of the limit surface and phase concentration may be expressed as
\begin{gather}\label{approx1}
\textbf{x}\cong\sum_{i=1}^{12}\textbf{x}_iN_i(s^1,s^2),\\
\label{approx2}\phi\cong\sum_{i=1}^{12}\phi_iN_i(s^1,s^2),
\end{gather}
where $\{N_i(s^1,s^2)\}$ are box-spline basis functions, whose exact forms are given in \cite{cirak2000subdivision}. Before proceeding, it is convenient to adopt the following abbreviated notation for partial derivatives:
\begin{equation*}
\partial_i(\bullet)\equiv\frac{\partial(\bullet)}{\partial \textbf{x}_i},\quad \hat{\partial}_i(\bullet)\equiv\frac{\partial(\bullet)}{\partial\phi_i}.
\end{equation*}

Let $N_e$ denotes the number of elements in the mesh. Substituting the finite element approximation (\ref{approx1}) and (\ref{approx2}) into the weak form (\ref{first_variation}) yields a set of nonlinear algebraic equations
\begin{subequations}
\begin{equation}\label{dPidrho}
\frac{\partial\Pi}{\partial\textbf{x}_i}=\sum_{k=1}^{N_e}\int_{\Omega_k}\big[\textbf{d}_\alpha\cdot(\partial_i\textbf{x})_{,\alpha}+\textbf{m}\cdot\partial_i\textbf{n}_{,\alpha}+\epsilon\sigma\partial_i e_g-\frac{p}{3}\textbf{n}\cdot\partial_i\textbf{x}\big]\sqrt{a}ds=0
\end{equation}
\begin{equation}\label{dPidphi}
\frac{\partial\Pi}{\partial \phi_i}=\sum_{k=1}^{N_e}\int_{\Omega_k}\big[\sigma\epsilon\hat{\partial}_i e_g+\big(\sigma W^{'}(\phi)+\lambda_\phi\big)\hat{\partial}_i\phi\big]\sqrt{a}ds=0
\end{equation}
\end{subequations}
the following identities can be shown to hold:
\begin{align*}
&\partial_i\textbf{n}_{,\alpha}=(\partial_i\textbf{n})_{,\alpha}=-N_{i,\beta\alpha}[\textbf{a}^\beta\otimes\textbf{n}]-N_{i,\beta}[\textbf{a}^\beta_{,\alpha}\otimes\textbf{n}]-N_{i,\beta}[\textbf{a}^\beta\otimes\textbf{n}_{,\alpha}],\\
&\partial_i e_g=-\frac{2e_g}{\sqrt{a}}\partial_i a+\frac{1}{2a}(\phi_{,1}^2\partial_i a_{22}+\phi_{,2}^2\partial_i a_{11}-2\phi_{,1}\phi_{,2}\partial_i a_{12}),\\
&\partial_i\textbf{x}=N_i\textbf{I}_3,\quad\partial_i a=2aN_{i,\alpha}\textbf{a}^\alpha\cdot\textbf{I}_3,\quad\partial_i a_{\alpha\beta}=N_{i,\alpha}\textbf{a}_\beta\cdot\textbf{I}_3+N_{i,\beta}\textbf{a}_\alpha\cdot\textbf{I}_3,\\
&\hat{\partial}_i e_g=\frac{a_{22}}{a}\phi_{,1}N_{i,1}+\frac{a_{11}}{a}\phi_{,2}N_{i,2}-\frac{a_{12}}{a}(\phi_{,1}N_{i,2}+\phi_{,2}N_{i,1}),
\end{align*}
where $\phi_{,j}=\sum_{i=1}^{12}\phi_i N_{i,j}(s^1,s^2)$.

Due to membrane fluidity characterizing the model, the energy functional (\ref{total_energy}) is invariant under any area-preserving diffeomophism of $\Sigma$ into itself - sometimes referred to as \quotes{reparametrization symmetry} in the literature. This, in turn, leads to massive ill-conditioning in (\ref{dPidrho}) and (\ref{dPidphi}), as pointed out, e.g., in \cite{ma2008viscous} and \cite{feng2006finite}. In order to overcome this difficulty, we employ the following radial-graph description
\begin{equation}\label{radiaGraph}
\textbf{x}=\rho(\textbf{X})\textbf{X},\quad\textrm{for all $\textbf{X}\in S^2$},
\end{equation}
where $S^2$ denotes the unit sphere and $\rho>0$ is a scalar field representing the magnitude of the radial position vector of the deformed surface. This effectively \quotes{mods out} the highly indeterminate in-plane deformation in consonance with the fact that the tangential balance of forces on $\Sigma$ is identically satisfied for this class of models, cf. \cite{healey2015symmetry}. From (\ref{radiaGraph}) the position vector and phase concentration corresponding to the $kth$ element (triangle $4-7-8$ in Figure \ref{ringPatch}) can now be approximated by
\begin{gather}
\label{approx1Element}\textbf{x}(s^1,s^2)\mid_{\Sigma_k}\cong\sum_{i=1}^{12}\rho_i\textbf{X}_iN_i(s^1,s^2),\\
\label{approx2Element}\phi(s^1,s^2)\mid_{\Sigma_k}\cong\sum_{i=1}^{12}\phi_iN_i(s^1,s^2),
\end{gather}
where $\textbf{X}_i$ is the fixed unit radial vector for node $i$ in the reference configuration $S^2$. We define the nodal variables for the $kth$ element via
\begin{equation}\label{unknownElement}
\textbf{u}^k=\big[\rho_1\quad\phi_1\mid\rho_2\quad\phi_2\mid\cdots\mid\rho_{12}\quad\phi_{12}\big],
\end{equation}
and we henceforth express the full vector of unknowns as
\begin{equation}\label{unknownFull}
\textbf{u}=\big[\rho_1\quad\cdots\quad\rho_n\quad\phi_1\quad\cdots\quad\phi_n\quad\lambda_s\quad\lambda_\phi\big]^T.
\end{equation}
where $n$ denotes the total number of nodes in the mesh. From (\ref{approx1}), (\ref{radiaGraph}) and (\ref{approx1Element}) we deduce $\frac{\partial\Pi}{\partial\rho_i}=\frac{\partial\Pi}{\partial\textbf{x}_i}\cdot\textbf{X}_i$. Hence the first variation of $\Pi$ has the following discrete form:
\begin{gather}
\label{F}\textbf{F}=\frac{\partial\Pi}{\partial\textbf{u}}=\bigg[\frac{\partial\Pi}{\partial\textbf{x}_1}\cdot\textbf{X}_1\quad\cdots\quad\frac{\partial\Pi}{\partial\textbf{x}_n}\cdot\textbf{X}_n\quad\frac{\partial\Pi}{\partial\phi_1}\quad\cdots\quad\frac{\partial\Pi}{\partial\phi_n}\quad\frac{\partial\Pi}{\partial\lambda_s}\quad\frac{\partial\Pi}{\partial\lambda_\phi}\bigg]^T,\\
\label{C}\textrm{with}\quad\frac{\partial\Pi}{\partial\lambda_s}=\sum_{k=1}^{NE}\int_{\Omega_k}\sqrt{a}ds-4\pi\quad\textrm{and}\quad\frac{\partial\Pi}{\partial\lambda_\phi}=\sum_{k=1}^{NE}\int_{\Omega_k}\phi\sqrt{a}ds-4\pi\mu,
\end{gather}
in practice we assemble $\textbf{F}$ and its gradient through element-wise numerical integration. Referring to (\ref{F}), we express the equilibrium equations as
\begin{equation}\label{f}
\textbf{F}(\textbf{u}, \lambda)=\textbf{0},
\end{equation}
where $\textbf{F}:\mathbb{R}^N\times\mathbb{R}\to\mathbb{R}^N$, with $N$ being the total number of unknowns. The generic symbol $\lambda\in R$ represents any of the parameter choices $\epsilon$, $B$, $\sigma$ or $\mu$.

\section{Symmetry Reduction}
\noindent The symmetries of the discrete equilibrium equations depend crucially upon that of the chosen mesh. It is shown in \cite{healey2015symmetry} that the Euler-Lagrange equations associated with (\ref{total_energy}) possess a plethora of equilibria classified by symmetry type according to specific subgroups of the orthogonal group $O(3)$. In particular, equilibrium configurations having the symmetries of the full icosahedral group $\mathbb{I}\oplus\mathbb{Z}^c_{2}$ are shown to exist in \cite{healey2015symmetry} as global solution branches bifurcating from the trivial symmetrical state $S^2$. Here $\mathbb{I}\oplus\mathbb{Z}^c_{2}$ refers the complete symmetry group of a regular icosahedron - comprising $60$ proper and $60$ improper rotations, cf. \cite{golubitsky2012singularities}. Motivated by both the \quotes{soccer-ball} equilibria observed in \cite{baumgart2003imaging} and the theoretical results of \cite{healey2015symmetry}, our goal here is to compute these solution branches. As such, we choose a mesh  having $\mathbb{I}\oplus\mathbb{Z}^c_{2}$ symmetry. For that purpose, we use a well-known algorithm for creating a geodesic sphere using a subdivided icosahedron, cf. \cite{sadourny1968integration, williamson1968integration}. In this work, our icosahedral-symmetric mesh is generated through five repeated subdivision of an icosahedron.

Equation (\ref{radiaGraph}) specifies the current position as a deformation of $S^2$, which we express as
\begin{equation}\label{deformation}
\textbf{f}(\textbf{X}):=\rho(\textbf{X})\textbf{X},\quad\textbf{f}:S^2\to\Sigma\subseteq\mathbb{R}^3.
\end{equation}

\noindent Let $\phi_m:=\phi\circ\textbf{f}$ denote the material version of the phase field on $S^2$. In Appendix B we demonstrate, using (\ref{deformation}), that the energy functional (\ref{total_energy}) is invariant under the transformations
\begin{equation}\label{action}
\rho(\textbf{X})\to\rho(\textbf{Q}^T\textbf{X}),\quad\phi_m(\textbf{X})\to\phi_m(\textbf{Q}^T\textbf{X})\quad\textrm{for all $\textbf{Q}\in O(3)$}.
\end{equation}

\noindent In particular, $\mathbb{I}\oplus\mathbb{Z}^c_{2}\subset O(3)$ is a subgroup. Presuming an $\mathbb{I}\oplus\mathbb{Z}^c_{2}$-symmetric mesh, denoted $S^2_d\subset\mathbb{R}^3$, that is $\textbf{Q}(S^2_d)=S^2_d$ for all $\textbf{Q}\in\mathbb{I}\oplus\mathbb{Z}^c_{2}$, then the action (\ref{action}) is inherited by the discrete field (\ref{unknownFull}) via matrix multiplication:
\begin{equation}\label{mapping}
\textbf{u}\to\mathcal{T}_Q\textbf{u}\quad\textrm{for all $\textbf{Q}\in\mathbb{I}\oplus\mathbb{Z}^c_{2}$},
\end{equation}
where $\textbf{Q}\mapsto\mathcal{T}_Q$ defines an $N\times N$ \textit{orthogonal matrix representation} of $\mathbb{I}\oplus\mathbb{Z}^c_{2}$, where $N$ denotes the total number of unknowns, cf. (\ref{f}). That is, $\mathcal{T}_{(\cdot)}$ is an $N\times N$ orthogonal matrix-valued function on $\mathbb{I}\oplus\mathbb{Z}^c_{2}$ satisfying:
\begin{enumerate}
\item $\mathcal{T}_{Q_1}\mathcal{T}_{Q_2}=\mathcal{T}_{Q_1Q_2}$ for all $\textbf{Q}_1, \textbf{Q}_2\in G$.
\item $\mathcal{T}^T_{Q}=\mathcal{T}_{Q^T}$ for all $\textbf{Q}\in G$, where $\mathcal{T}^T_Q$ denotes transpose of the matrix $\mathcal{T}_Q$.
\item $\mathcal{T}_I=\mathcal{I}$, where $\mathcal{I}$ is the $N\times N$ identity matrix.
\end{enumerate}

\noindent Finally, the invariance of the potential energy functional under (\ref{action}) and the inherited discrete action (\ref{mapping}) together imply
\begin{theorem}\label{equivariance}
The discrete energy function, denoted $\hat{\Pi}$, arising from (\ref{energy_ref}), (\ref{approx1Element}) and (\ref{approx2Element}), is invariant under (\ref{mapping}), viz.,
\begin{equation}\label{energy_invariance}
\hat{\Pi}(\mathcal{T}_Q\textbf{u},\lambda)=\hat{\Pi}(\textbf{u},\lambda)\quad\textrm{for all $\textbf{Q}\in\mathbb{I}\oplus\mathbb{Z}^c_{2}$}.
\end{equation}
Moreover, the discrete equilibrium equations (\ref{f}) are \textit{equivariant}, viz.,
\begin{equation}\label{equiv_relation}
\textbf{F}(\mathcal{T}_Q\textbf{u},\lambda)=\mathcal{T}_Q\textbf{F}(\textbf{u},\lambda)\quad\textrm{for all $\textbf{Q}\in\mathbb{I}\oplus\mathbb{Z}^c_{2}$}.
\end{equation}
\end{theorem}

\noindent We note that direct differentiation of (\ref{energy_invariance}) yields $\mathcal{T}_Q^T\hat{\Pi}_\textbf{u}(\mathcal{T}_Q\textbf{u},\lambda)=\hat{\Pi}_\textbf{u}(\textbf{u},\lambda):=\textbf{F}(\textbf{u},\lambda)$ for all $\textbf{Q}\in\mathbb{I}\oplus\mathbb{Z}^c_{2}$. But the orthogonality of $\mathcal{T}_Q$, viz., $\mathcal{T}_Q^T=\mathcal{T}_Q^{-1}$ gives (\ref{equiv_relation}). A detailed proof is presented in Appendix B.

In order to exploit \textit{equivariance}, we define the fixed-point space
\begin{equation}\label{fixed_point}
\mathcal{V}_G=\big{\{}\textbf{u}\in R^{N}:\mathcal{T}_{Q}\textbf{u}=\textbf{u}\quad\textrm{for all $\textbf{Q}\in G:=\mathbb{I}\oplus\mathbb{Z}^{c}_{2}$}\big{\}},
\end{equation}
which is readily shown to be a subspace. Combining (\ref{equiv_relation}) and (\ref{fixed_point}), it follows that
\begin{equation}
\mathcal{T}_Q\textbf{F}(\textbf{u},\lambda)=\textbf{F}(\mathcal{T}_Q\textbf{u},\lambda)=\textbf{F}(\textbf{u},\lambda)\quad\textrm{for all $\textbf{Q}\in\mathbb{I}\oplus\mathbb{Z}_2^c$ and $(\textbf{u},\lambda)\in\mathcal{V}_G\times\mathbb{R}$},
\end{equation}
In other words, the nonlinear map $\textbf{u}\mapsto\textbf{F}(\textbf{u},\lambda)$ has the linear invariant subspace $\mathcal{V}_G$. From group representation theory \cite{miller1972symmetry}, the orthogonal projection operator is readily obtained as
\begin{equation}
\mathcal{P}=\frac{1}{NG}\sum_{i=1}^{NG}\mathcal{T}_{Q_i},
\end{equation}
and the dimension of $\mathcal{V}_G$ is given by
\begin{equation}
dim \mathcal{V}_G=\frac{1}{NG}\sum_{i=1}^{NG}tr\mathcal{T}_{Q_i}
\end{equation}
where $NG$ is the order of the subgroup $G$, $\textbf{Q}_i$ range over all its elements. In appendix A, we present a novel algorithm to compute $\mathcal{P}$ and $dim \mathcal{V}_G$ given a mesh possessing $\mathbb{I}\oplus\mathbb{Z}^c_{2}$ symmetry.

The $G$-reduced problem $\textbf{F}_G$ is defined by projecting $\textbf{F}$ onto the fixed point space $\mathcal{V}_G\times\mathbb{R}$
\begin{equation}\label{reduced_prob}
\textbf{F}_G(\textbf{u}, \lambda)\equiv\mathcal{P}\circ\textbf{F}(\textbf{u},\lambda)=\textbf{0},
\end{equation}
where $\textbf{u}\in\mathcal{V}_G$. Note that equation (\ref{reduced_prob}) represents a dimensional reduction in problem size. The significance of the reduced problem is summarized in the following important theorem \cite{healey1988group}, which follows directly from (\ref{equiv_relation}):
\begin{theorem}\label{solutionEquiv}
A point $(\textbf{u}_0,\lambda_0)\in\mathcal{V}_G\times\mathbb{R}$ is a solution to equation (\ref{f}) if and only if it's also a solution to the $G$-reduced problem.
\end{theorem}
This ensures that the solutions of the lower-dimensional problem (\ref{reduced_prob}) provide exact $G$-symmetric solutions to the full problem (\ref{f}). In appendix C we provides a scheme for reducing the discretized system for a given symmetry group $G$.

\section{Computational Results}
As mentioned previously, cf. (\ref{f}), our problem depends upon various parameters. Accordingly, we employ numerical continuation or path following in order to explore families of equilibria. We consistently do so by continuing in one of the parameters with the others fixed. Details for numerical continuation can be found in \cite{keller1987lectures}. For fixed internal pressure $p\geq 0$, the results of \cite{healey2015symmetry} yield the existence of bifurcating branches of icosahedral-symmetric solutions from the spherical state if $\mu$ and $\epsilon$ satisfy the \textit{characteristic equation}
\begin{equation}\label{charac}
\kappa:=\frac{1}{\epsilon}=-\frac{l(l+1)}{W^{''}(\mu)}\quad\textrm{for modes $l=6, 10, 12, 15, 16, 18, \cdots$}
\end{equation}
where $W^{''}(\mu)>0$, which is called \textit{spinodal region} of the double-well potential. Icosahedral-symmetric solution branches corresponding to modes $l=6, 10$ are explored in this paper.

\begin{figure}[!htb]
    \centering
    \begin{subfigure}[b]{0.45\textwidth}
    \includegraphics[width=\textwidth]{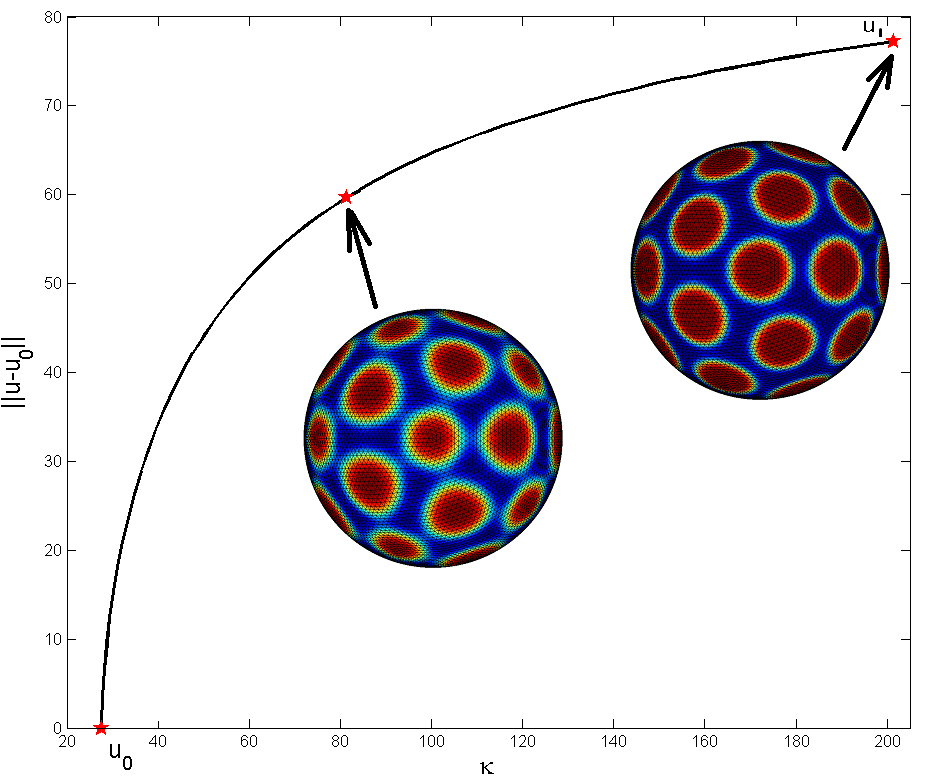}
    \caption{$\kappa$-solution branch}
    \label{Kappa_l10_mu0}
    \end{subfigure}
    \begin{subfigure}[b]{0.475\textwidth}
    \includegraphics[width=\textwidth]{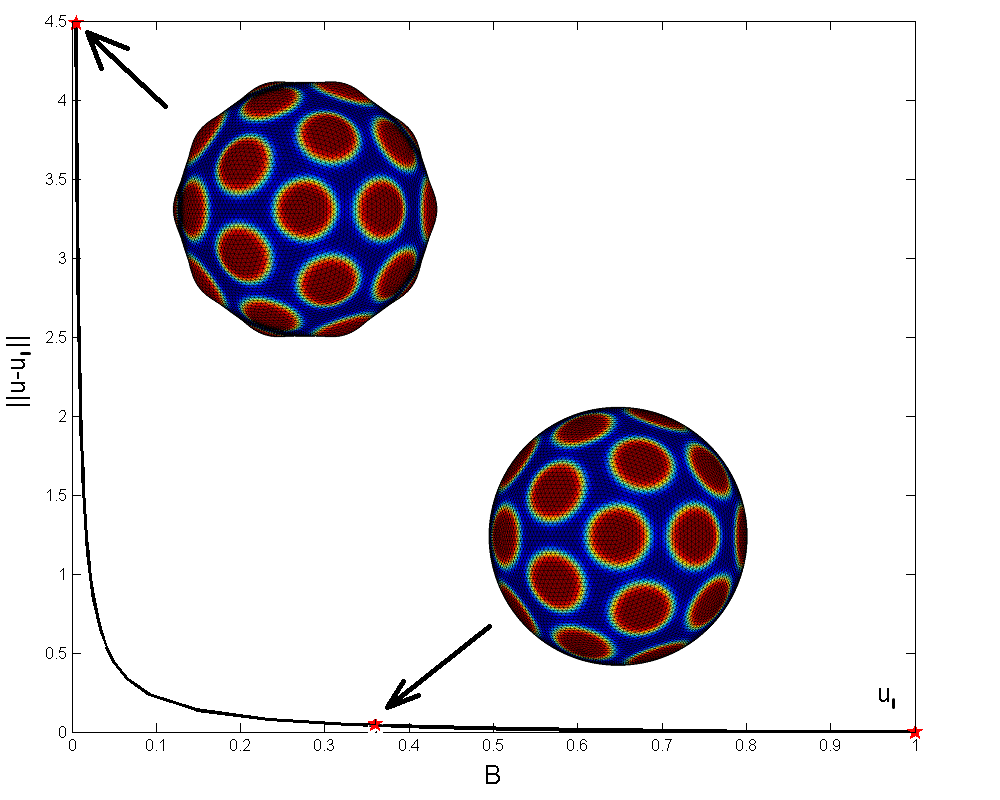}
    \caption{$B$-solution branch}
    \label{B_l10_mu0}
    \end{subfigure}
    \caption{Solution branches for mode $l=10$ with $p=1$, $\mu=0$ and $\sigma=1$.}
\end{figure}

\begin{figure}[!htb]
    \centering
    \includegraphics[width=0.45\textwidth]{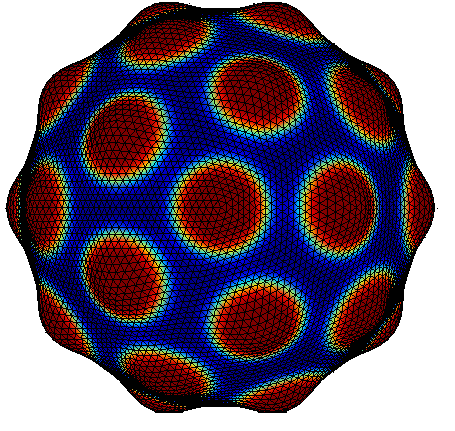}
    \caption{Vesicle for mode $l=10$ with $\sigma=1.2$, $p=1$, $\mu=0$, $\kappa=200$ and $B=0.005$}
    \label{Deformation_l10_mu1p2_B0p005}
\end{figure}

As discussed in the previous section, we generate an icosahedral-symmetric mesh via five repeated subdivisions of an icosahedron. This results in $n=10242$ mesh points leading to $N=2n+2=20486$ degrees of freedom in (\ref{f}). Nonetheless, the reduction method of Section 4 delivers a $G$-reduced problem (\ref{reduced_prob}) with merely $410$ unknowns. We start by imitating the strategy used in \cite{healey2013wrinkling}, \cite{healey2007two}, \cite{sanjaythesis}, employing $\kappa:=\frac{1}{\epsilon}$ as a bifurcation/continuation parameter. Figure \ref{Kappa_l10_mu0} shows the non-trivial $\kappa$-solution branch for mode $l=10$ with $\mu=0$, $p=1$, $B=1$ and $\sigma=1$. Note the considerable sharpening of the interface as $\kappa$ is increased (decreasing of $\epsilon$). At this stage, there is no noticeable deformation, which suggests that the bending stiffness $B=1$ is relatively large. As such, we fix $\kappa:=\frac{1}{\epsilon}=200$ and restart continuation by decreasing $B$ as the continuation parameter. Figure \ref{B_l10_mu0} shows the corresponding $B$-solution branch. We now observe a noticeable deformation as we decrease the bending stiffness $B$. This strategy was first proposed in \cite{sanjaythesis} in order to solve the axisymmetric case for this model. The parameter $\sigma$ balances the curvature and phase contributions to the total energy (\ref{energy}). At this stage, we fix $B=0.005$ and $\kappa=200$ and employ $\sigma$ as the continuation parameter, Figure \ref{Deformation_l10_mu1p2_B0p005} shows one solution point on the $\sigma$-solution branch with $\sigma=1.2$. We now see that the deformation is less spherical when $\sigma$ is increased, the shape resembles the experimental observation from \cite{baumgart2003imaging}. We observed convergence difficulties when attempting to increase $\sigma$, while decreasing $\sigma$ caused no difficulties.

\begin{figure}[!htb]
    \centering
    \begin{subfigure}[b]{0.32\textwidth}
    \includegraphics[width=\textwidth]{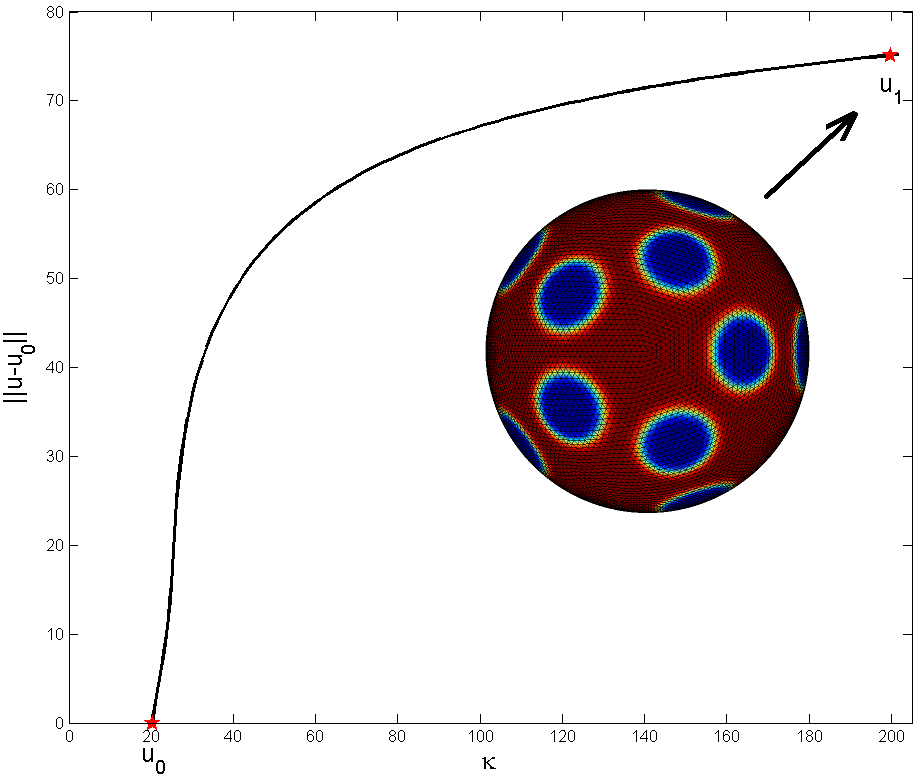}
    \caption{$\mu=0.4$}
    \end{subfigure}
    \begin{subfigure}[b]{0.32\textwidth}
    \includegraphics[width=\textwidth]{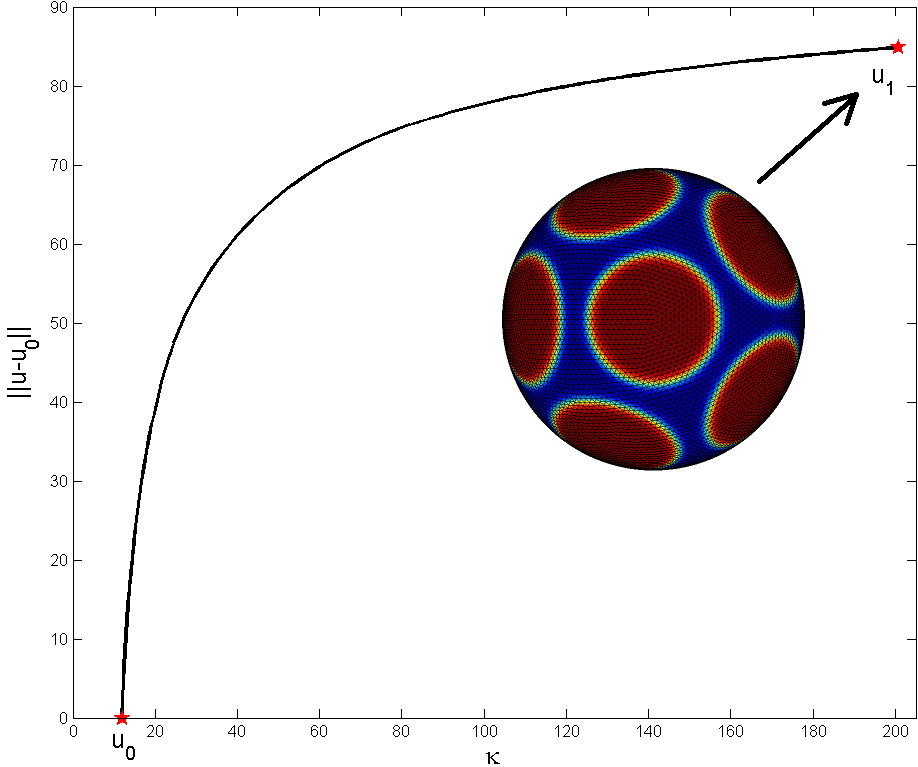}
    \caption{$\mu=0.2$}
    \end{subfigure}
    \begin{subfigure}[b]{0.32\textwidth}
    \includegraphics[width=\textwidth]{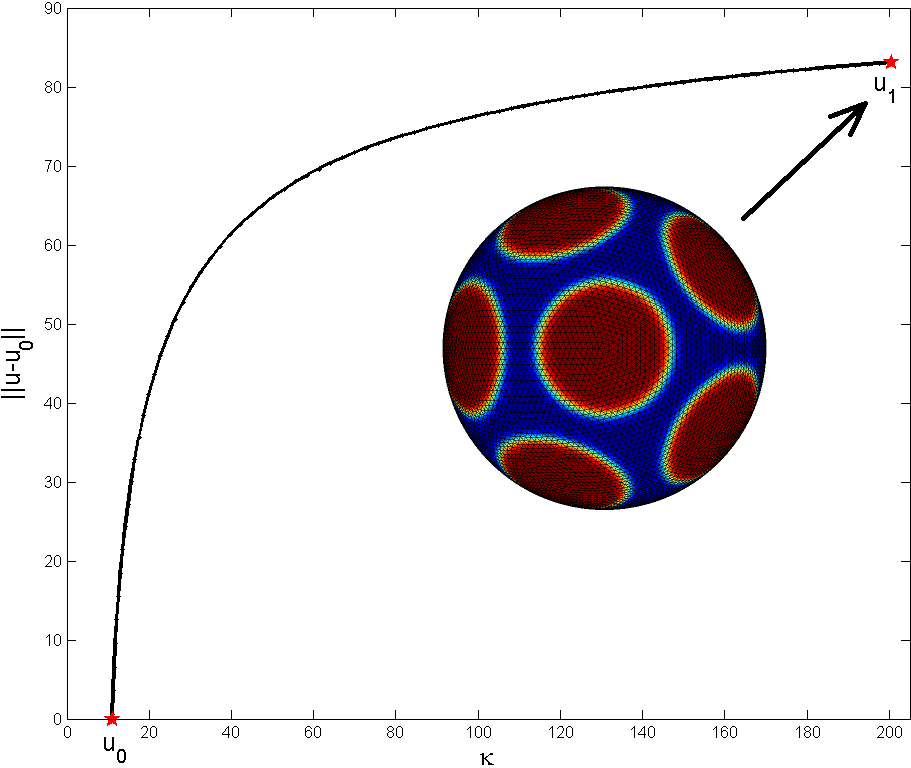}
    \caption{$\mu=0.1$}
    \end{subfigure}

    \begin{subfigure}[b]{0.32\textwidth}
    \includegraphics[width=\textwidth]{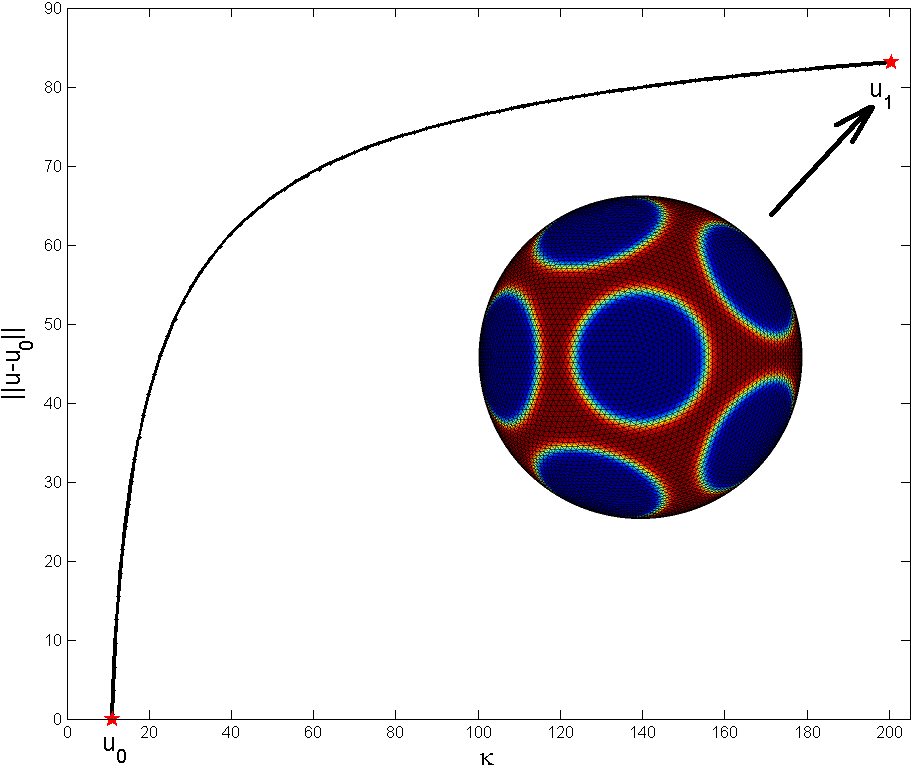}
    \caption{$\mu=-0.1$}
    \end{subfigure}
    \begin{subfigure}[b]{0.32\textwidth}
    \includegraphics[width=\textwidth]{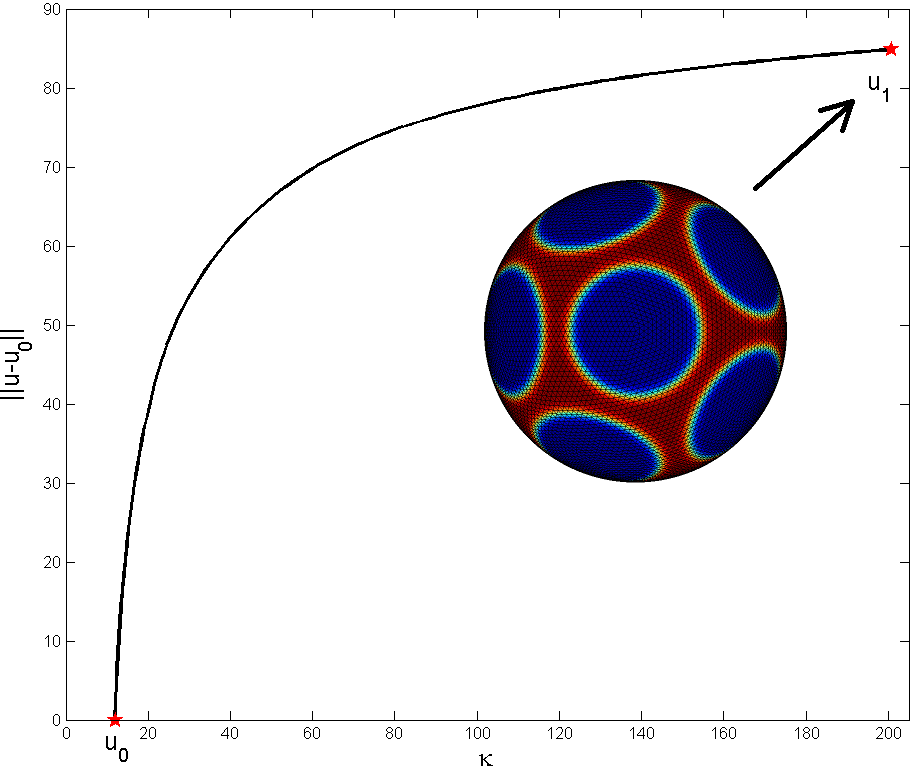}
    \caption{$\mu=-0.2$}
    \end{subfigure}
    \begin{subfigure}[b]{0.32\textwidth}
    \includegraphics[width=\textwidth]{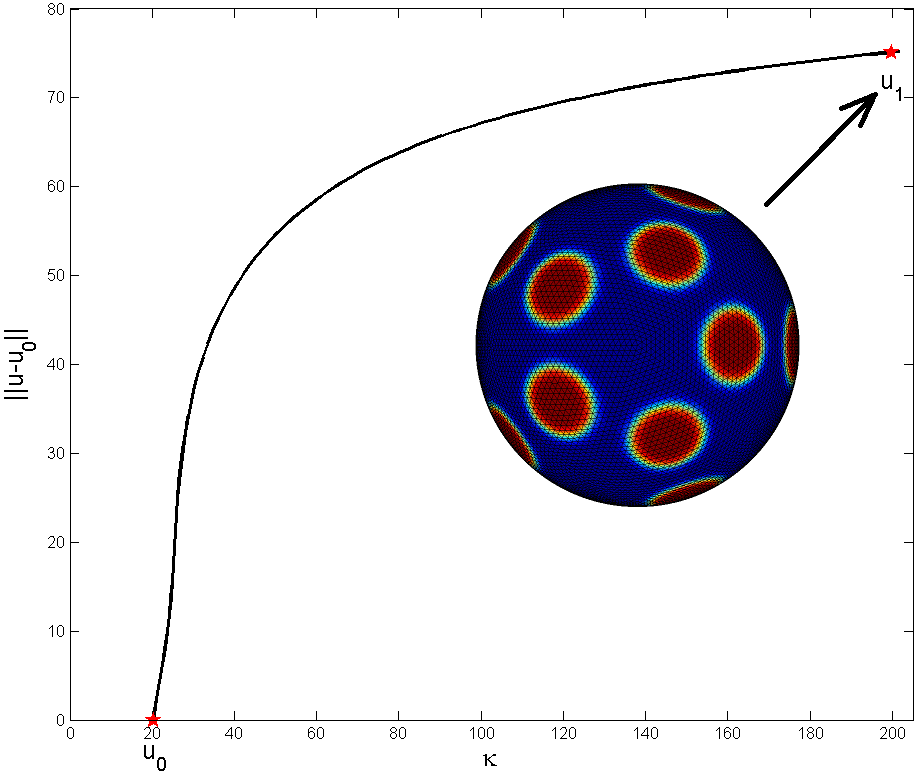}
    \caption{$\mu=-0.4$}
    \end{subfigure}
    \caption{$\kappa$-solution branches for mode $l=6$ with $B=1$, $p=1$ and $\sigma=1$.}
    \label{Kappa_l6}
\end{figure}

\begin{figure}[!htb]
    \centering
    \begin{subfigure}[b]{0.31\textwidth}
    \includegraphics[width=\textwidth]{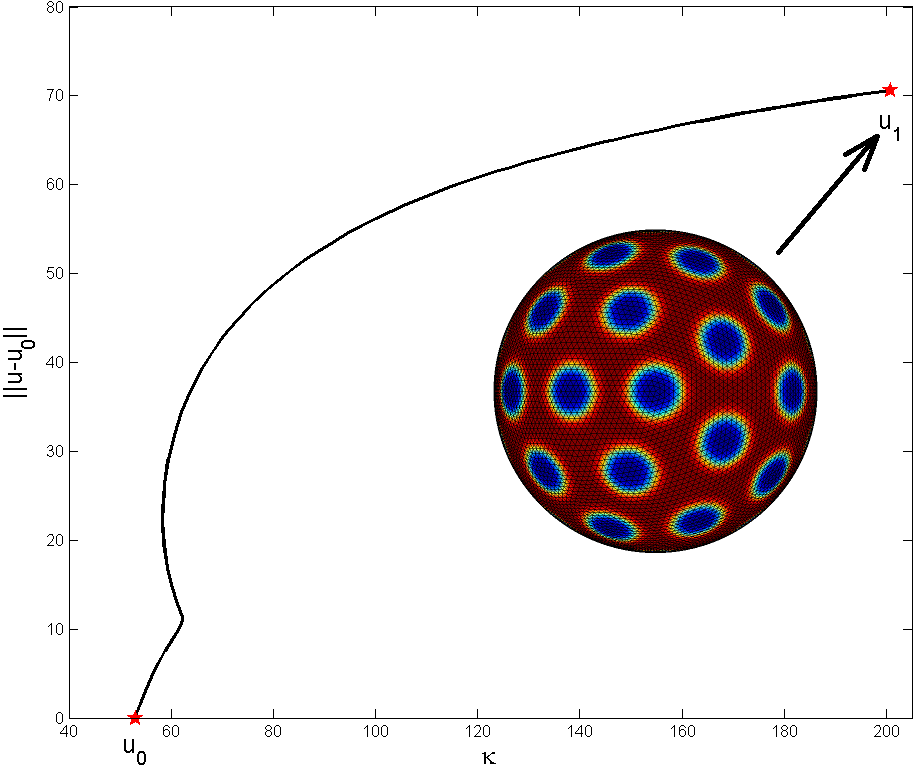}
    \caption{$\mu=0.4$}
    \label{Kappa_l10_mu0p4}
    \end{subfigure}
    \begin{subfigure}[b]{0.32\textwidth}
    \includegraphics[width=\textwidth]{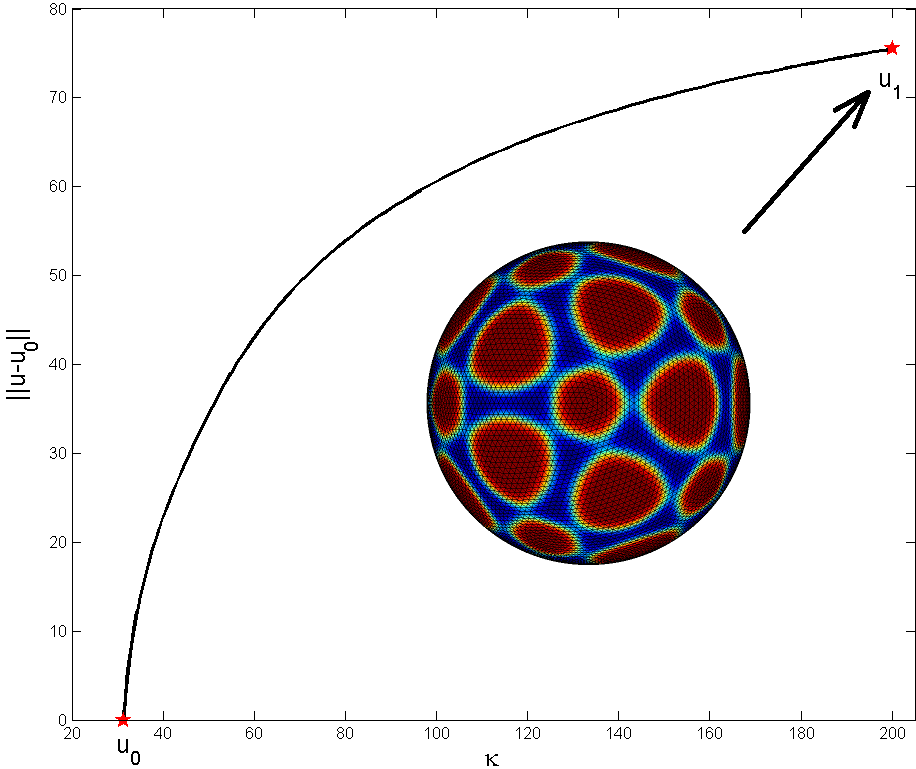}
    \caption{$\mu=0.2$}
    \end{subfigure}
    \begin{subfigure}[b]{0.32\textwidth}
    \includegraphics[width=\textwidth]{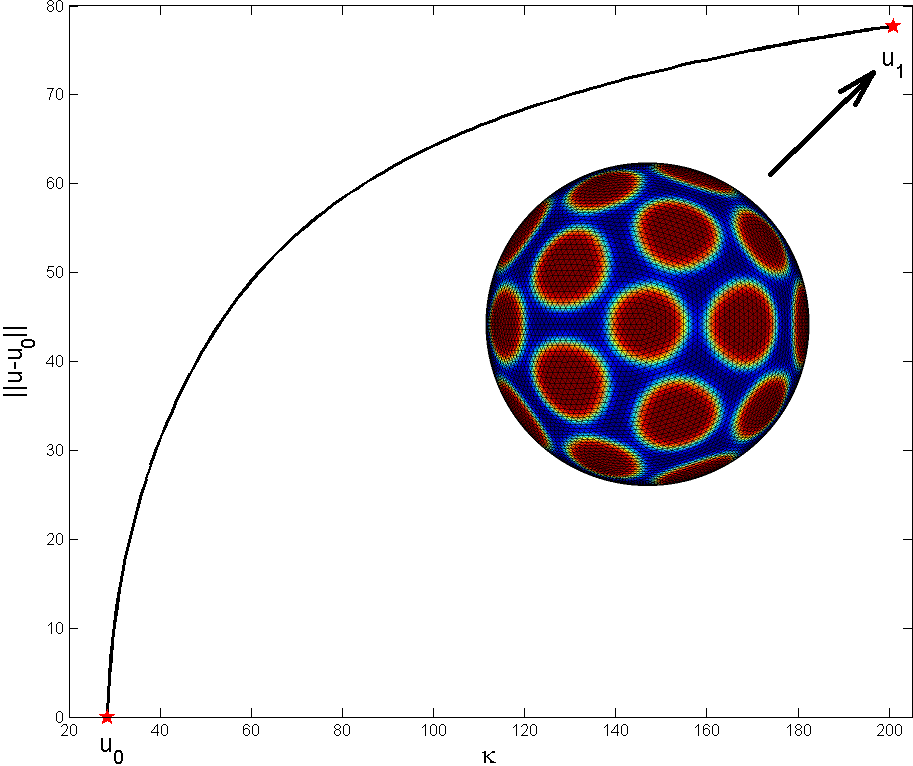}
    \caption{$\mu=0.1$}
    \end{subfigure}

    \begin{subfigure}[b]{0.32\textwidth}
    \includegraphics[width=\textwidth]{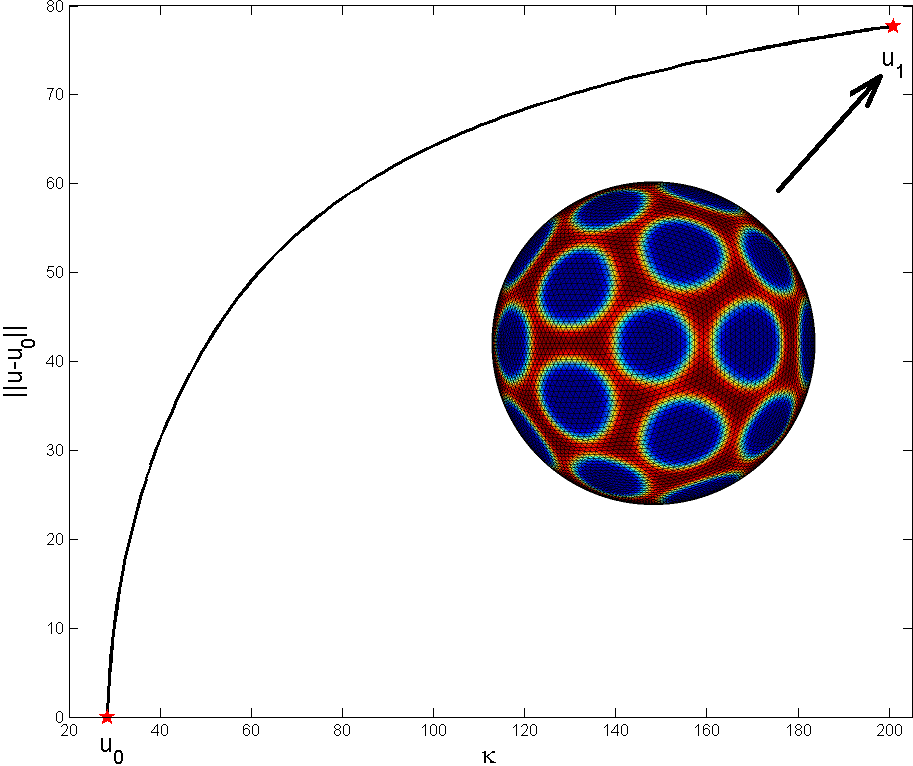}
    \caption{$\mu=-0.1$}
    \end{subfigure}
    \begin{subfigure}[b]{0.32\textwidth}
    \includegraphics[width=\textwidth]{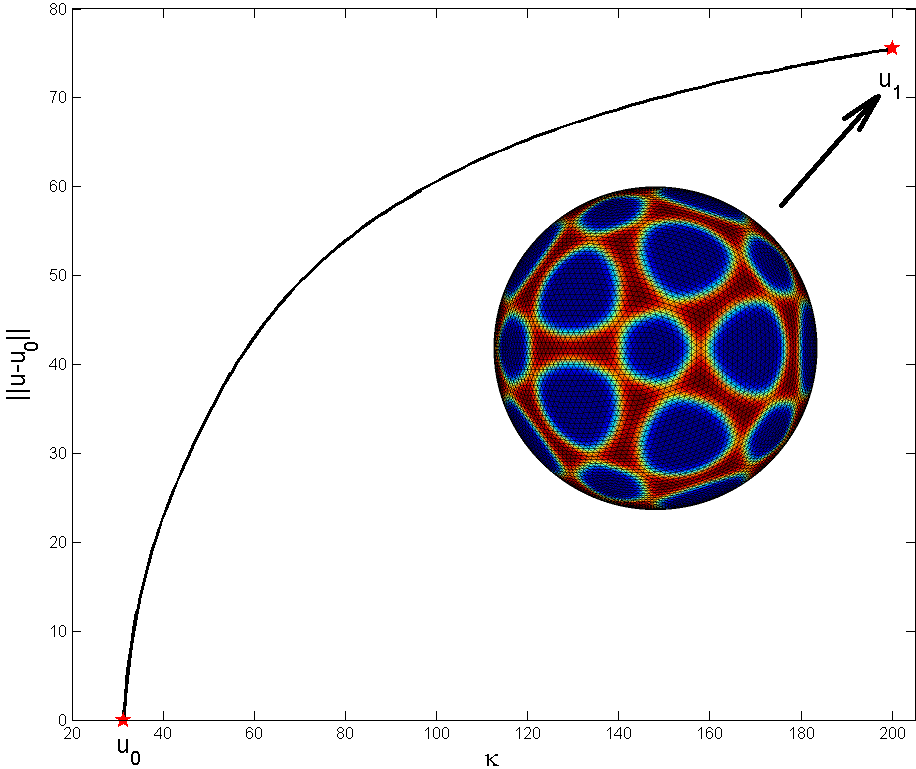}
    \caption{$\mu=-0.2$}
    \end{subfigure}
    \begin{subfigure}[b]{0.31\textwidth}
    \includegraphics[width=\textwidth]{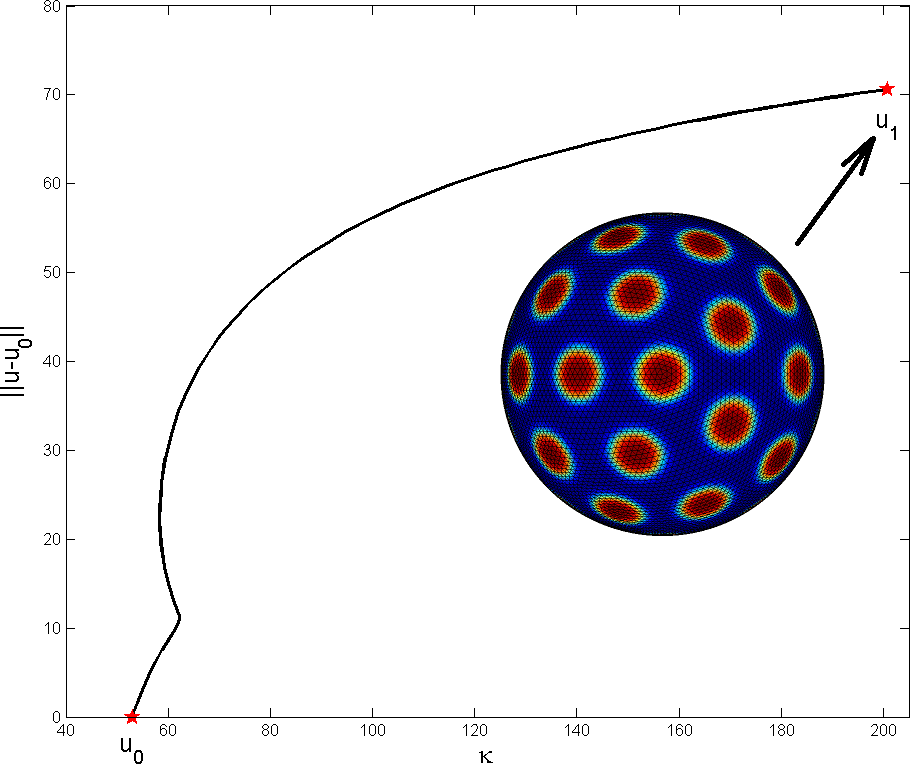}
    \caption{$\mu=-0.4$}
    \label{Kappa_l10_mun0p4}
    \end{subfigure}
    \caption{$\kappa$-solution branches for mode $l=10$ with $B=1$, $p=1$ and $\sigma=1$.}
    \label{Kappa_l10}
\end{figure}

\begin{figure}[!htb]
    \centering
    \begin{subfigure}[b]{0.321\textwidth}
    \includegraphics[width=\textwidth]{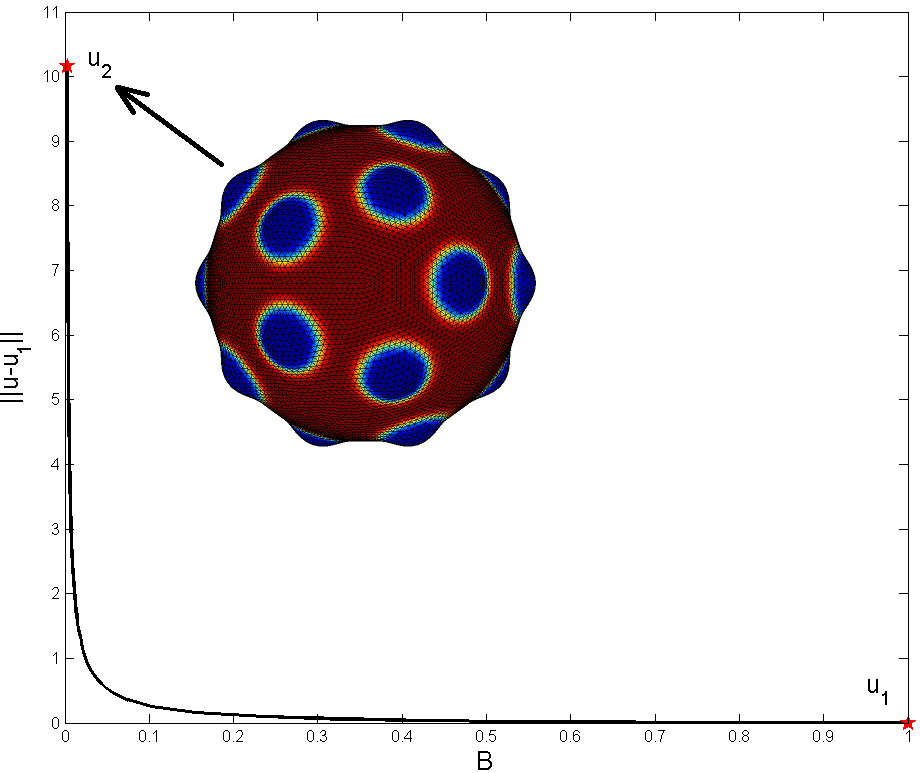}
    \caption{$\mu=0.4$}
    \end{subfigure}
    \begin{subfigure}[b]{0.32\textwidth}
    \includegraphics[width=\textwidth]{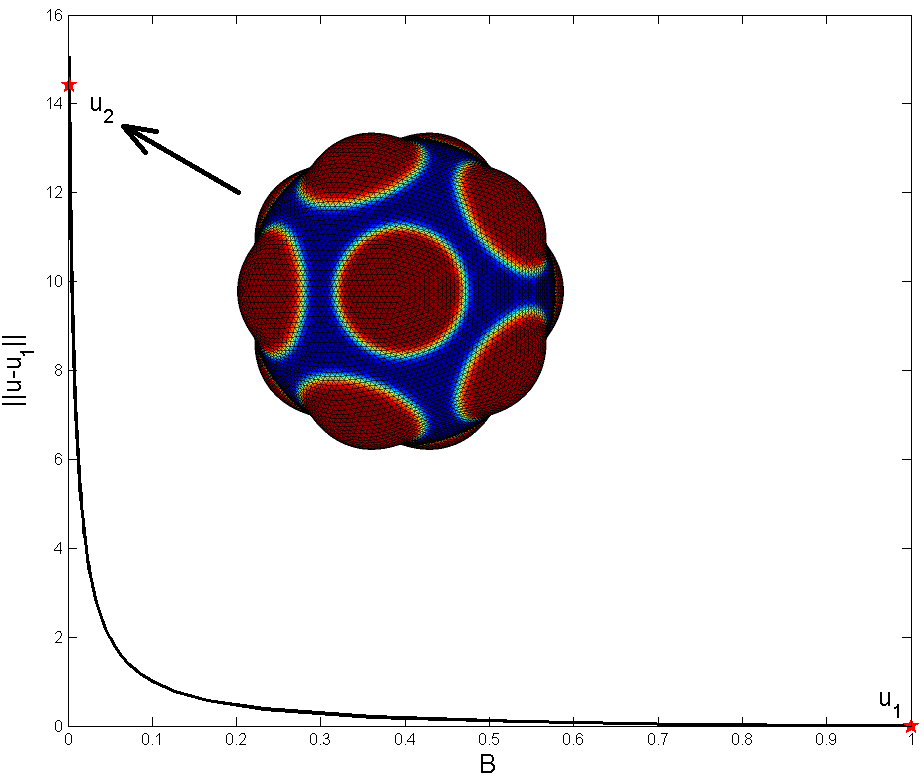}
    \caption{$\mu=0.2$}
    \end{subfigure}
    \begin{subfigure}[b]{0.32\textwidth}
    \includegraphics[width=\textwidth]{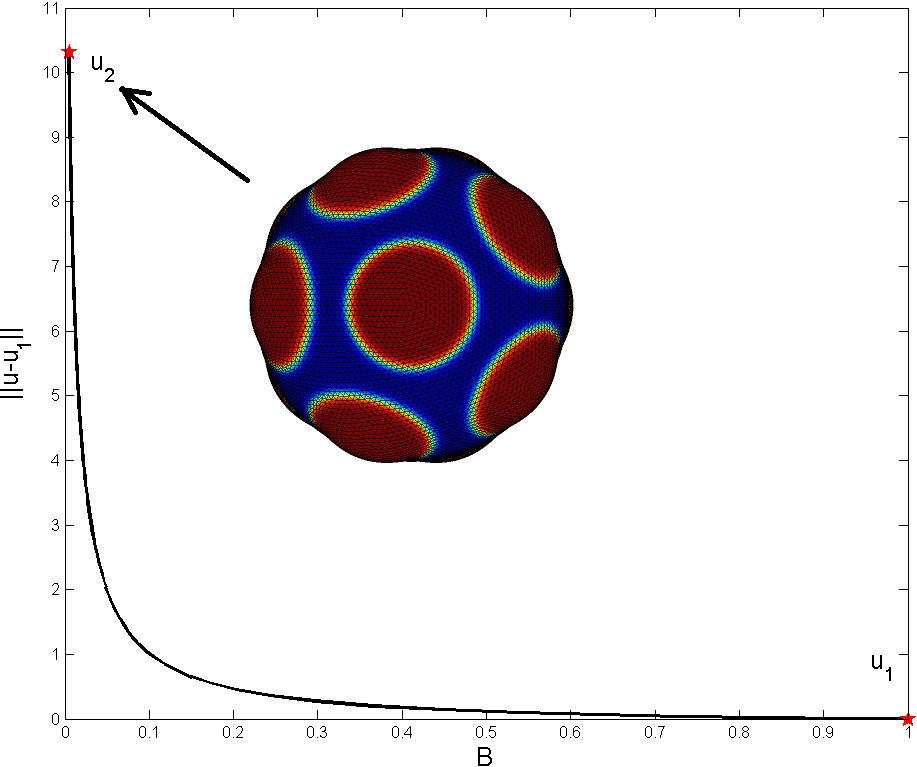}
    \caption{$\mu=0.1$}
    \end{subfigure}

    \begin{subfigure}[b]{0.32\textwidth}
    \includegraphics[width=\textwidth]{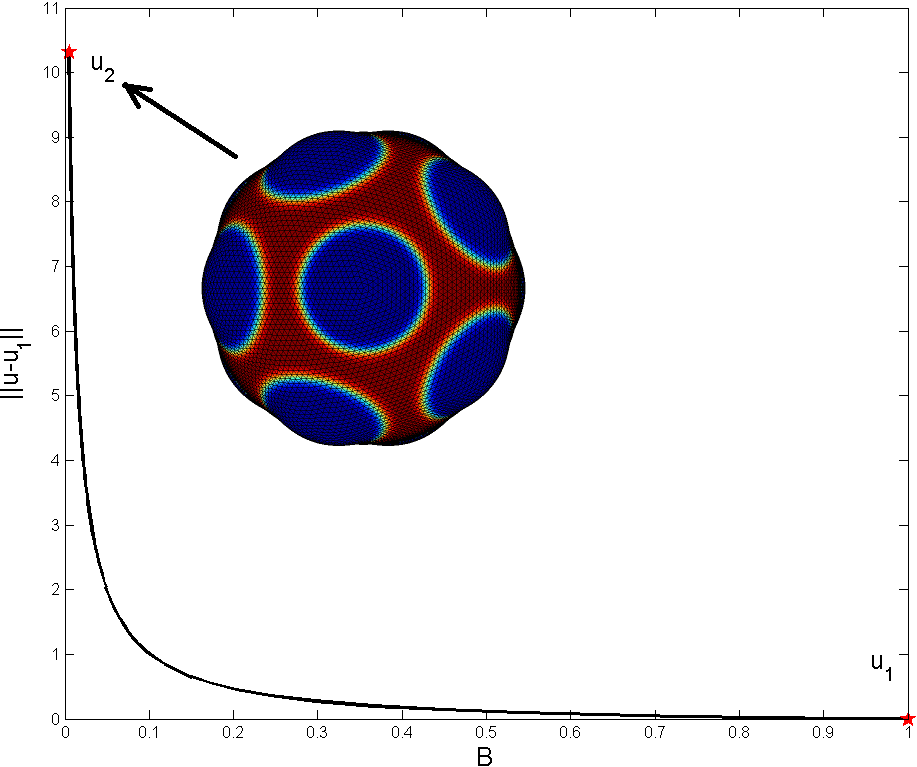}
    \caption{$\mu=-0.1$}
    \end{subfigure}
    \begin{subfigure}[b]{0.32\textwidth}
    \includegraphics[width=\textwidth]{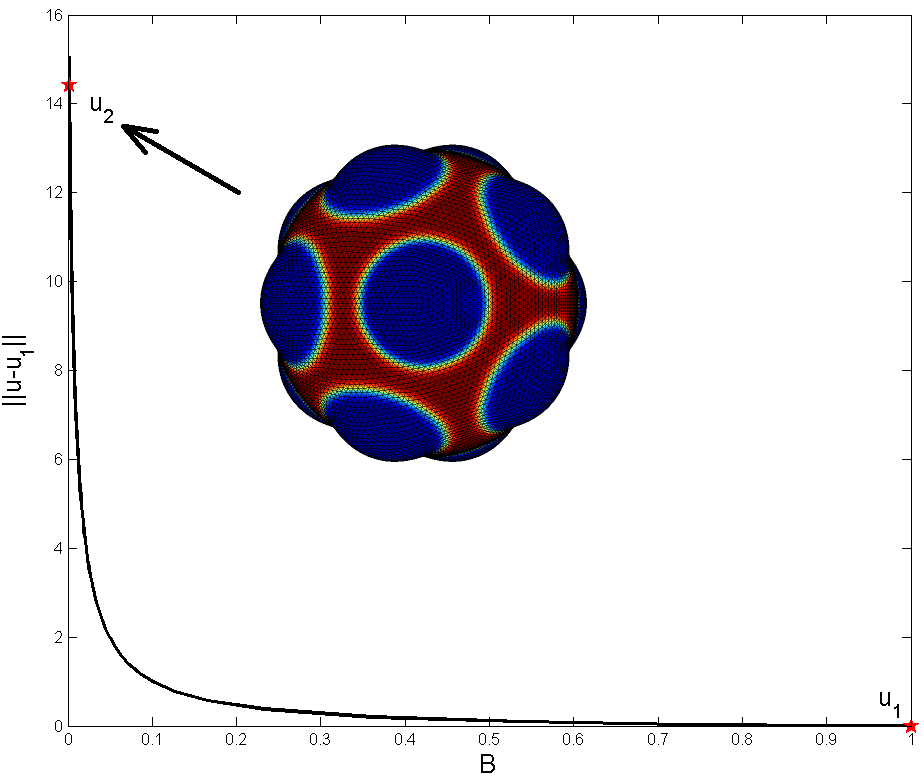}
    \caption{$\mu=-0.2$}
    \end{subfigure}
    \begin{subfigure}[b]{0.321\textwidth}
    \includegraphics[width=\textwidth]{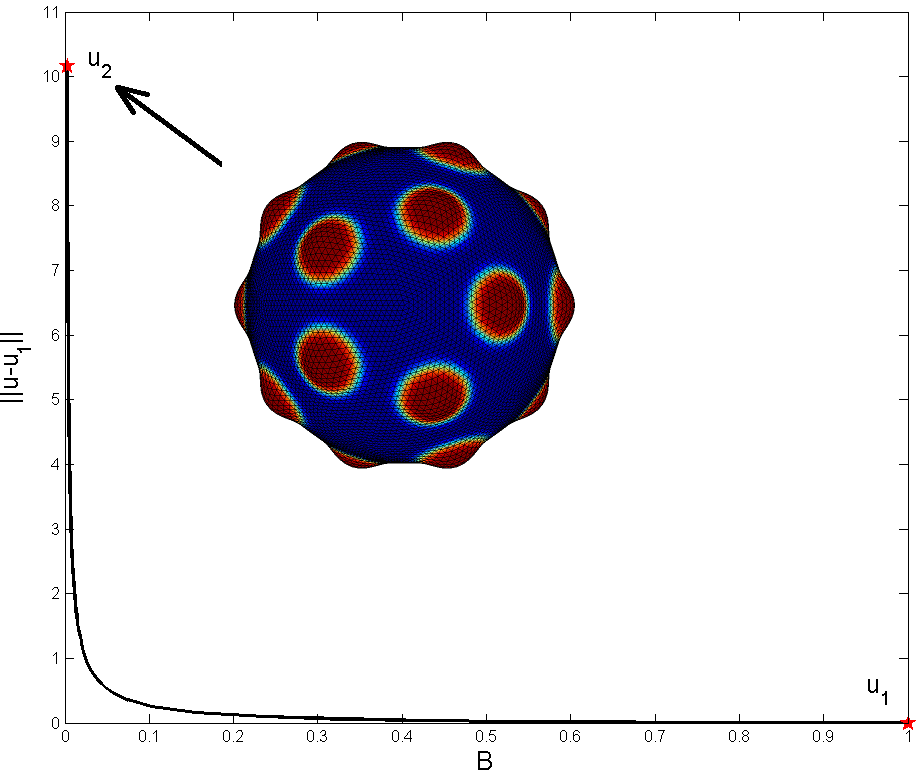}
    \caption{$\mu=-0.4$}
    \end{subfigure}
    \caption{$B$-solution branches for mode $l=6$ with $\kappa=200$, $p=1$ and $\sigma=1$.}
    \label{B_kappa200_l6}
\end{figure}

\begin{figure}[!htb]
    \centering
    \begin{subfigure}[b]{0.325\textwidth}
    \includegraphics[width=\textwidth]{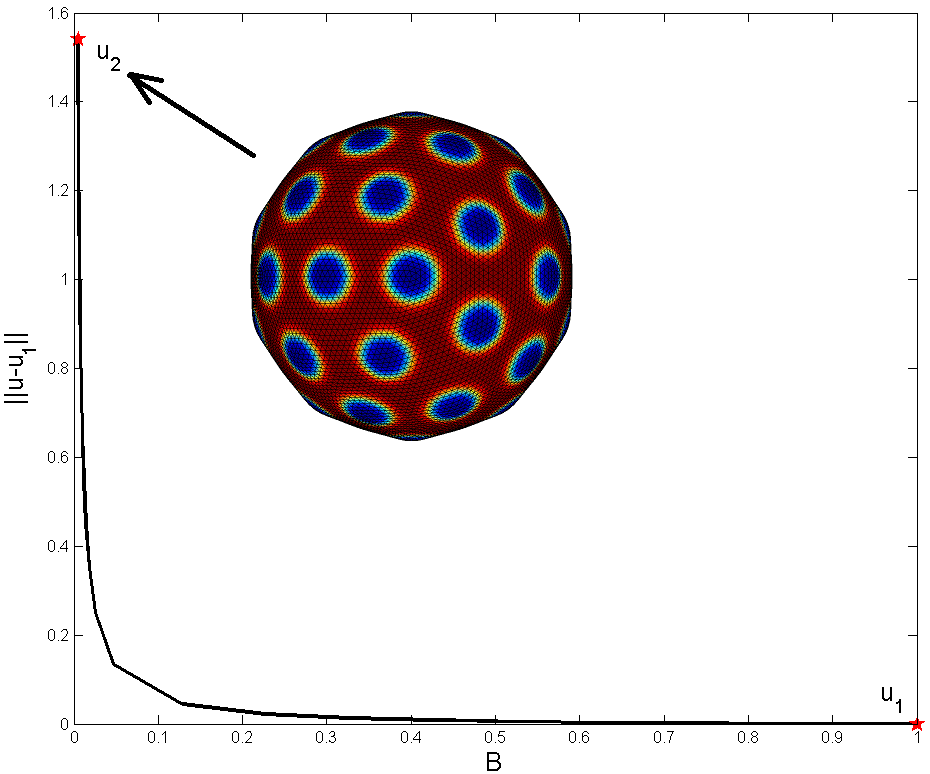}
    \caption{$\mu=0.4$}
    \label{B_kappa200_l10_mu0p4}
    \end{subfigure}
    \begin{subfigure}[b]{0.32\textwidth}
    \includegraphics[width=\textwidth]{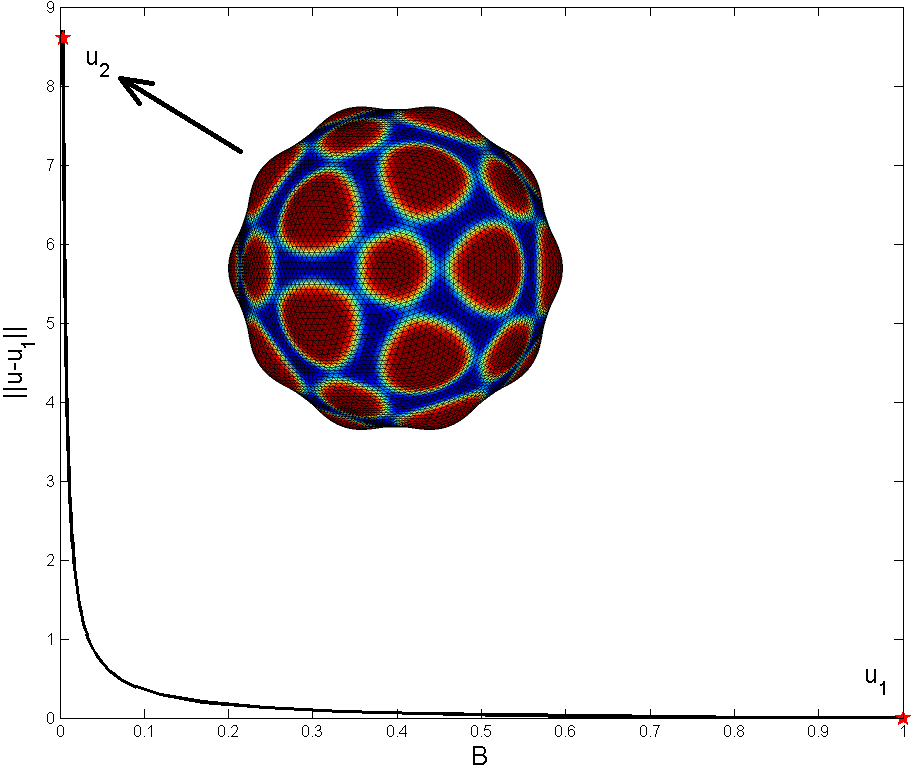}
    \caption{$\mu=0.2$}
    \end{subfigure}
    \begin{subfigure}[b]{0.32\textwidth}
    \includegraphics[width=\textwidth]{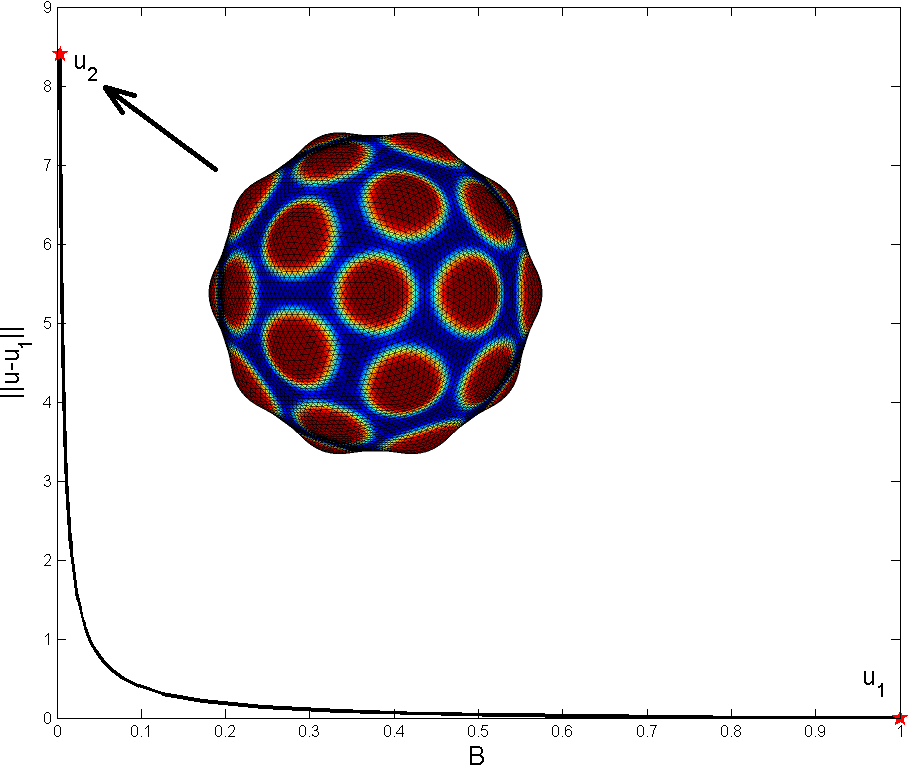}
    \caption{$\mu=0.1$}
    \end{subfigure}

    \begin{subfigure}[b]{0.32\textwidth}
    \includegraphics[width=\textwidth]{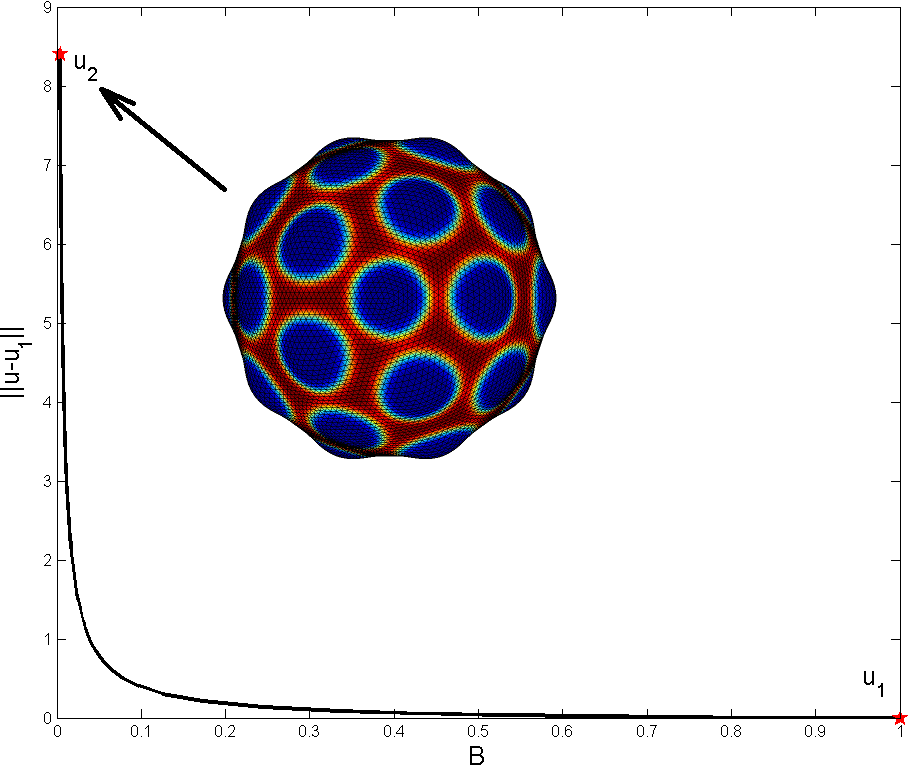}
    \caption{$\mu=-0.1$}
    \end{subfigure}
    \begin{subfigure}[b]{0.32\textwidth}
    \includegraphics[width=\textwidth]{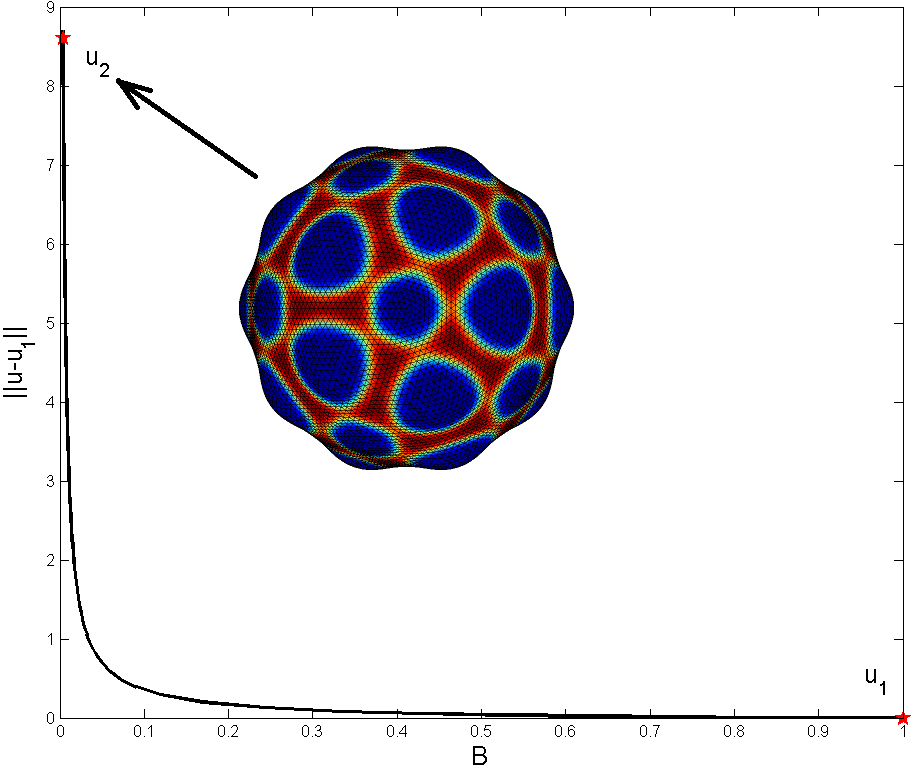}
    \caption{$\mu=-0.2$}
    \end{subfigure}
    \begin{subfigure}[b]{0.325\textwidth}
    \includegraphics[width=\textwidth]{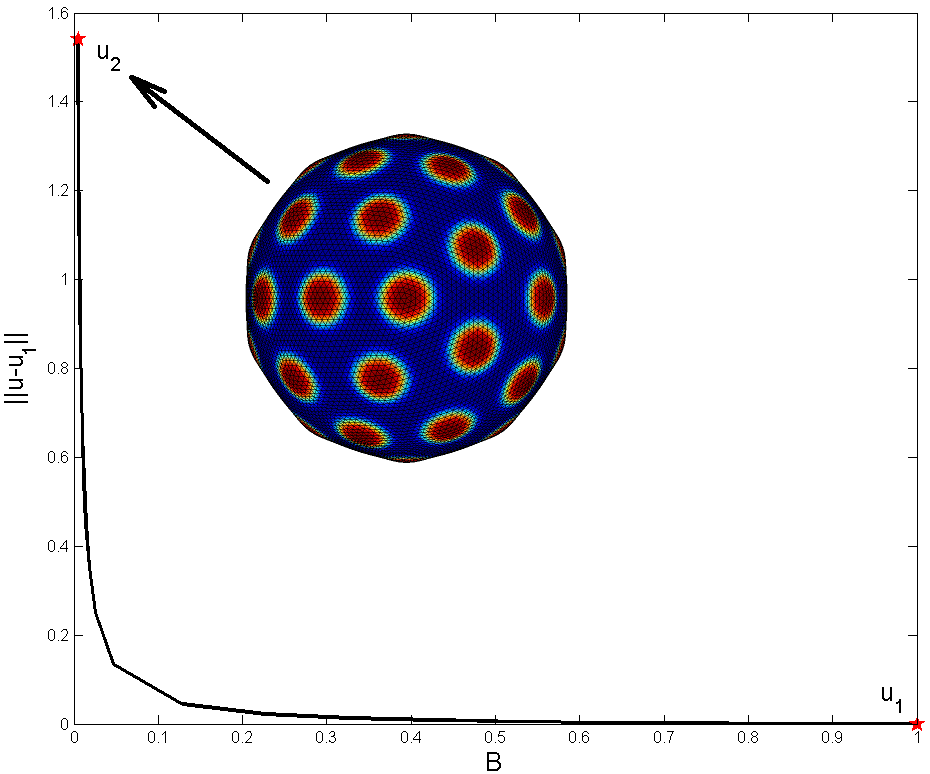}
    \caption{$\mu=-0.4$}
    \label{B_kappa200_l10_mun0p4}
    \end{subfigure}
    \caption{$B$-solution branches for mode $l=10$ with $\kappa=200$, $p=1$ and $\sigma=1$.}
    \label{B_kappa200_l10}
\end{figure}

Another interesting parameter is $\mu$, which represents the average phase concentration on the surface. Figures \ref{Kappa_l6} and \ref{Kappa_l10} show the $\kappa$-solution branches for modes $l=6,10$ with various $\mu$ values. As can be observed, $\mu$ controls the phase-pattern on the surface. Note that there are two \textit{limit points} in Figures \ref{Kappa_l10_mu0p4} and \ref{Kappa_l10_mun0p4}. Following the aforementioned strategy, Figures \ref{B_kappa200_l6} and \ref{B_kappa200_l10} show the corresponding $B$-solution branches for modes $l=6,10$, after which we fix $\kappa=200$ and decrease the bending stiffness $B$. Again, we can observe a noticeable deformation as we decrease $B$. The results summarized in Figures \ref{Kappa_l6}-\ref{B_kappa200_l10} are obtained by initializing $\mu$ to various values and computing the branches as before.


\section{Concluding Remarks}
We present here a systematic approach to computing two-phase equilibria of lipid-bilayer structures with icosahedral symmetry. We uncover a wide variety of such configurations as the various parameters in the model are varied via numerical continuation. Our approach can be easily modified to account for any of the multitude of symmetry types, the existence of solutions for which is obtained in \cite{healey2015symmetry}. It's enough to choose a mesh having the symmetry of one of the cataloged subgroups in \cite{golubitsky2012singularities}, cf. XIII, Theorem 9.9. We can then use the methodology of Section 4 tailored to that particular subgroup.

Our results do not address the stability of the numerous equilibria found. Indeed, a check for local stability (local energy minimum) involves identifying the definiteness of the complete tangent stiffness matrix at a given equilibrium (the second-derivative test) - not just the relatively small tangent stiffness associated with the reduced problem (\ref{reduced_prob}). For example, with the mesh employed in obtaining the results of Section 5, the Jacobian for (\ref{F}) and (\ref{C}) at an equilibrium is a $20486\times 20486$ matrix, which is  unsymmetric due to the constraints (\ref{C}). The method of \cite{kumar2010generalized} can be used to obtain the complete symmetric tangent matrix, now of size $20484\times 20484$. But the latter is not banded and hence of formidable size. Further block diagonalization of the tangent stiffness via group representation theory, along the lines of \cite{wohlever1995group} (but a much simpler symmetry) can be obtained. This would seem to be required here in order to reliably address local stability of equilibria. In this respect, numerical gradient-flow methods \cite{elliott2013computation}, \cite{funkhouser2010dynamics}, \cite{taniguchi1996shape} have an apparent advantage of tending to stable solutions. However, parameter values must be juggled in such a way that the particular equilibrium sought renders the potential energy a global minimum - or at least it should be in a deep energy well. This overlooks the possibility of meta-stable states. For example, it is not clear that all configurations observed in \cite{baumgart2003imaging} are global energy minima.

\appendix
\section{Algorithm to compute $\mathcal{T}$ for a given mesh under $\mathbb{I}\oplus\mathbb{Z}^c_2$}
\begin{figure}[!htb]
    \centering
    \includegraphics[width=0.6\textwidth]{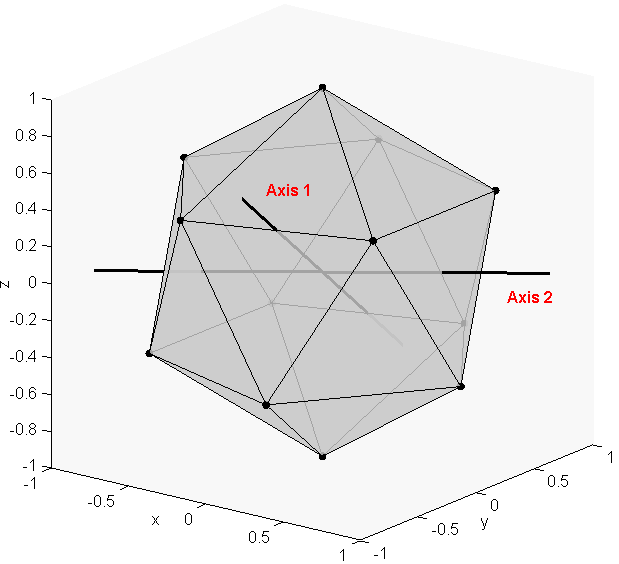}
    \caption{Illustration of an icosahedron}
    \label{icosa}
\end{figure}

\noindent In this section, we present an algorithm to compute a faithful $N\times N$ orthogonal matrix representation of the icosahedral group $\mathbb{I}\oplus\mathbb{Z}^c_2$,
which comprises all $60$ proper rotations of an icosahedron into itself (denoted by $\mathbb{I}$) plus all improper rotations resulting from multiplying each element in $\mathbb{I}$ by negative identity. An icosahedron is shown in Figure \ref{icosa}. All the $60$ proper rotations are accounted for as follows: $\frac{1\times30}{2}$ rotations about the axis connecting $30$ edges centers, $\frac{2\times20}{2}$ rotations about the axis connecting $20$ face centers and $\frac{4\times12}{2}$ plus one identity, a total of $60$ rotations, formally characterized as $\mathbb{I}=\dot{\bigcup}\quad ^{6}\mathbb{Z}_5\quad\dot{\bigcup}\quad ^{10}\mathbb{Z}_3\quad\dot{\bigcup}\quad ^{15}\mathbb{Z}_2$ in \cite{golubitsky2012singularities}.

Let's define two rotational matrices w.r.t. Axis 1 and Axis 2 in Figure \ref{icosa},
\begin{gather*}
\textbf{C}=\begin{pmatrix}
  t_ce_1^2+\cos\theta_c & t_ce_1e_2-\sin\theta_ce_3 & t_ce_1e_3+\sin\theta_ce_2 \\
  t_ce_1e_2+\sin\theta_c e_3 & t_ce_2^2+\cos\theta_c & t_ce_2e_3-\sin\theta_ce_1 \\
  t_ce_1e_3-\sin\theta_c e_2 & t_ce_2e_3+\sin\theta_ce_1 & t_ce_3^2+\cos\theta_c
\end{pmatrix},\\
\textrm{with $\theta_c=\pi$ and $t_c=1-\cos\theta_c$},
\end{gather*}
\begin{gather*}
\textbf{D}=\begin{pmatrix}
  t_df_1^2+\cos\theta_d & t_df_1f_2-\sin\theta_df_3 & t_df_1f_3+\sin\theta_df_2 \\
  t_df_1f_2+\sin\theta_de_3 & t_df_2^2+\cos\theta_d & t_df_2f_3-\sin\theta_df_1 \\
  t_df_1f_3-\sin\theta_de_2 & t_df_2f_3+\sin\theta_df_1 & t_df_3^2+\cos\theta_d
\end{pmatrix},\\
\textrm{with $\theta_d=\frac{2\pi}{3}$ and $t_d=1-\cos\theta_d$},
\end{gather*}
where $(e_1,e_2,e_3)$ and $(f_1,f_2,f_3)$ are Axis 1 and Axis 2 respectively.

All $60$ rotations can be realized by compositions of rotations $\textbf{C}$ and $\textbf{D}$. For example, $\textbf{CD}$ represent a rotation of $\frac{6\pi}{5}$ about the $z$-axis. We refer \cite{simingthesis} for a full list of $60$ rotations. Multiplying the negative $3\times 3$ identity matrix on the $60$ proper rotations gives us additional $60$ improper reflections, a total of $120$ elements in the group $\mathbb{I}\oplus\mathbb{Z}^c_2$. We summarize our scheme for computing the projection operator in Algorithm \ref{Alg_ProjOper}.
\begin{algorithm}
\caption{Algorithm to compute the projector operator $\mathcal{P}$}
\label{Alg_ProjOper}
\begin{algorithmic}[1]
\State Generate a $n$-point icosahedral-symmetric mesh, let the $3\times n$ matrix $X$ to represent the position of all nodal points, set $mG=dim\mathcal{V}_G$.
\State Initialize $\mathcal{P}\gets\textbf{0}_{2n\times 2n}$ and $mG\gets 0$
\State Loop $i:=1:120$:
\State \qquad$\textbf{T}\gets\textbf{0}_{2n\times 2n}$
\State \qquad$\textbf{X}_R\gets\textbf{R}_i\textbf{X}$ locations of the mesh points after rotation $\textbf{R}_i$
\State\qquad Loop $j:=1:n$:
\State\qquad\qquad $k\gets$ index $k$ such that $jth$ column of $\textbf{X}_R$ $\equiv$ $kth$ column of $\textbf{X}$
\State\qquad\qquad$\textbf{T}[j,k]\gets 1$ and $\textbf{T}[n+j,n+k]\gets 1$
\State\qquad$\mathcal{P}\gets\mathcal{P}+\textbf{T}$
\State\qquad$mG\gets mG+trace(\textbf{T})$
\State $\mathcal{P}\gets\frac{\mathcal{P}}{120}$
\State $mG\gets\frac{mG}{120}$
\end{algorithmic}
\end{algorithm}

The table below lists the dimension of the reduced problem ($dim \mathcal{V}_G$) for different number of mesh points. As can be easily observed, there is a significant reduction of dimension in the reduced problem.
\begin{center}
  \begin{tabular}{ | r | r | r | }
    \hline
    number of subdivision of the icosahedron & number of mesh points & $dim \mathcal{V}_G$ \\ \hline \hline
    2 & 162 & 8 \\ \hline
    3 & 642 & 20 \\ \hline
    4 & 2562 & 60 \\ \hline
    5 & 10242 & 204 \\
    \hline
  \end{tabular}\label{dim_red}
\end{center}

\section{Proof of Theorem \ref{equivariance}}
We first define some terminology before giving the proof. Let $\textbf{X}\in S^2$ denote positions on the unit sphere, $S^2$. Consider a smooth map $\textbf{f}:S^2\to\mathbb{R}^3$, such that $\textbf{x}=\textbf{f}(\textbf{X})$ is the position vector on the deformed surface $\Sigma:=\textbf{f}(S^2)$. Let $\textbf{X}=\textbf{X}(S^1,S^2)$, where $(s^1,s^2)$ are curvilinear coordinates. We employ the convected description
\begin{equation}\label{des}
\textbf{x}=\textbf{f}(\textbf{X}(s^1,s^2)):=\textbf{x}(s^1,s^2).
\end{equation}
The covariant basis vector fields are then given by
\begin{equation}\label{covbasis}
\textbf{A}_\alpha=\textbf{X}_{,\alpha}\quad\textrm{and}\quad\textbf{a}_\alpha=\textbf{x}_{,\alpha}=\textbf{F}\textbf{A}_\alpha,
\end{equation}
where $\textbf{F}:=\nabla\textbf{f}$ denotes the deformation gradient. We have the following convenient representation for the deformation gradient and its restricted inverse, as follows:
\begin{equation}\label{FFinv}
\textbf{F}=\textbf{a}_\alpha\otimes\textbf{A}^\alpha\quad\textrm{and}\quad\textbf{F}^{-1}=\textbf{A}_\alpha\otimes\textbf{a}^\alpha.
\end{equation}

We first demonstrate that the total potential energy (\ref{total_energy}) is invariant under a certain representation of the full orthogonal group. In order to carry this out, we first need to express the former in terms of a material description relative to $S^2$. The material description of phase field is $\phi_m:=\phi\circ\textbf{f}$. The chain rule delivers
\begin{equation}\label{grad}
\nabla_\Sigma\phi=\textbf{F}^{-T}\nabla\phi_m.
\end{equation}
The material version of the mean curvature is a bit more involved. We first write the outward unit normal field to $\Sigma$ in the material description, viz., $\textbf{n}_m=\textbf{n}\circ\textbf{f}$. The chain rule now implies
\begin{equation}
\nabla_\Sigma\textbf{n}=\nabla\textbf{n}_m\textbf{F}^{-1}.
\end{equation}
We note that $\textbf{n}_m$ is the unique vector field satisfying
\begin{equation}\label{Fnm}
\textbf{F}^T\textbf{n}_m=\textbf{0}\quad\textrm{on $S^2$}.
\end{equation}
Recalling that $H=-tr(\nabla_\Sigma\textbf{n})/2$, we then have
\begin{equation}\label{meancurv}
H_m=-tr(\nabla\textbf{n}_m\textbf{F}^{-1})/2.
\end{equation}
With (\ref{grad}) and (\ref{meancurv}) in hand, the change of variables formula and (\ref{total_energy}) yield
\begin{multline}\label{energy_ref2}
\Pi[(\textbf{f},\phi_m)]=\int_{S^2}\bigg(BH_m^2+\sigma\big(\frac{\epsilon}{2}\parallel\textbf{F}^{-T}\nabla\phi_m\parallel^2+W(\phi)\big)\bigg)Jds+\lambda_s(\int_{S^2}Jds-4\pi)\\
+\lambda_\phi(\int_{S^2}\phi_mJds-4\pi\mu)-\frac{p}{3}\int_{S^2}\textbf{f}\cdot\textbf{n}_mJds,
\end{multline}
where $J=\sqrt{\textbf{F}^T\textbf{F}}$ is the area ratio. It's convenient to introduce $\boldsymbol\Phi:=(\textbf{f},\phi_m)$, in which case the left side of (\ref{energy_ref2}) is simply denoted $\Pi[\boldsymbol\Phi]$ The first variation condition for (\ref{energy_ref2}) delivers the weak form of the equilibrium equations:
\begin{equation}\label{variationDef}
\frac{d}{d\alpha}\Pi[\boldsymbol\Phi+\alpha\textbf{Y}]\mid_{\alpha=0}=\big<\delta\Pi[\boldsymbol\Phi],\textbf{Y}\big>=0,
\end{equation}
for all admissible variations $\textbf{Y}=(\mathbf{h},\psi)$.

Given an orthogonal transformation $\textbf{Q}\in O(3)$, we define its action via
\begin{equation}
(\textbf{f},\phi_m)\to\mathcal{T}_Q(\textbf{f},\phi_m):=(\textbf{Q}\textbf{f}(\textbf{Q}^T\textbf{X}),\phi_m(\textbf{Q}^T\textbf{X})),
\end{equation}
or simply
\begin{equation}\label{orthtransform}
\boldsymbol\Phi\to\mathcal{T}_Q\boldsymbol\Phi.
\end{equation}
For a fixed orthogonal transformation $\textbf{Q}$, observe that $\mathcal{T}_Q$ is a linear operator on pairs $\textbf{u}=(\textbf{f},\phi_m)$. We now claim that the energy functional (\ref{energy_ref2}) is invariant under (\ref{orthtransform}), viz.,
\begin{equation}\label{PiTF}
\Pi(\mathcal{T}_Q\boldsymbol\Phi)=\Pi(\boldsymbol\Phi)\quad\textrm{for all $\textbf{Q}\in O(3)$}.
\end{equation}
To see this, observe that (\ref{orthtransform}) and the chain rule lead to
\begin{equation}\label{FgradPhi}
\textbf{F}(\textbf{X})\to\textbf{Q}\textbf{F}(\textbf{Q}^T\textbf{X})\textbf{Q}^T,\quad\nabla\phi_m(\textbf{X})\to\textbf{Q}\nabla\phi_m(\textbf{Q}^T\textbf{X}).
\end{equation}
We the immediately find that
\begin{gather}\label{gradJmap}
\parallel\textbf{F}^{-T}(\textbf{X})\nabla\phi_m(\textbf{X})\parallel^2\to\parallel\textbf{F}^{-T}(\textbf{Q}^T\textbf{X})\nabla\phi_m(\textbf{Q}^T\textbf{X})\parallel^2 ,\\
\label{Jmap}J(\textbf{X})\to J(\textbf{Q}^T\textbf{X}).
\end{gather}
Moreover, by virtue of (\ref{Fnm}) and (\ref{FgradPhi}), we deduce
\begin{equation}\label{ntoQn}
\textbf{n}_m(\textbf{X})\to\textbf{Q}\textbf{n}_m(\textbf{Q}^T\textbf{X}).
\end{equation}
Indeed, observe from (\ref{Fnm}) and (\ref{ntoQn}) that
\begin{equation*}
\textbf{F}^T(\textbf{X})\textbf{n}_m(\textbf{X})\to\textbf{Q}\textbf{F}^T(\textbf{Q}^T\textbf{X})\textbf{Q}^T\textbf{Q}\textbf{n}_m(\textbf{Q}^T\textbf{X})=\textbf{Q}\textbf{F}^T(\textbf{Q}^T\textbf{X})\textbf{n}_m(\textbf{Q}^T\textbf{X})=\textbf{0}.
\end{equation*}
Next observe from (\ref{orthtransform}) and (\ref{ntoQn}) that
\begin{equation}\label{nmap}
\textbf{f}(\textbf{X})\cdot\textbf{n}_m(\textbf{X})\to\textbf{f}(\textbf{Q}^T\textbf{X})\cdot\textbf{n}_m(\textbf{Q}^T\textbf{Q}^T).
\end{equation}
The relations (\ref{meancurv}), (\ref{FgradPhi}) and (\ref{ntoQn}) now yield
\begin{equation}\label{Hmap}
H_m(\textbf{X})\to H_m(\textbf{Q}^T\textbf{X}).
\end{equation}
Finally, (\ref{gradJmap}), (\ref{Jmap}), (\ref{nmap}) and (\ref{Hmap}) show that the entire integrand on the left side of (\ref{PiTF}) is a function of the independent variable $\textbf{Y}=\textbf{Q}^T\textbf{X}$. Since $\textbf{Q}\in O(3)$, the invariance (\ref{PiTF}) then follows from the change of variables formula.

Next we take an arbitrary directional derivative of both sides of (\ref{PiTF}), viz.,
\begin{equation*}
\frac{d}{d\alpha}\Pi(\mathcal{T}_Q(\boldsymbol\Phi+\alpha\textbf{Y}))\mid_{\alpha=0}=\frac{d}{d\alpha}\Pi(\boldsymbol\Phi+\alpha\textbf{Y})\mid_{\alpha=0},
\end{equation*}
leading to
\begin{equation}\label{firstVariationDot}
\big<\delta\Pi(\mathcal{T}_Q\boldsymbol\Phi),\mathcal{T}_Q\textbf{Y}\big>=\big<\delta\Pi(\boldsymbol\Phi),\textbf{Y}\big>,
\end{equation}
for all $\textbf{Q}\in O(3)$ and admissible variations $\textbf{Y}$. The left side of (\ref{firstVariationDot}) can be expressed as
\begin{equation}\label{firstVariationDot2}
\big<\delta\Pi(\mathcal{T}_Q\boldsymbol\Phi),\textbf{Y}\big>=\big<\delta\Pi(\boldsymbol\Phi),\mathcal{T}_{Q^T}\textbf{Y}\big>=\big<\mathcal{T}^{*}_{Q^T}\delta\Pi(\boldsymbol\Phi),\textbf{Y}\big>,
\end{equation}
where $\mathcal{T}_Q^*$ denotes the adjoint operator of $\mathcal{T}_Q$. It's easy to show $\mathcal{T}^*_{Q}=\mathcal{T}_{Q^T}$ for all $\textbf{Q}\in O(3)$. This along with (\ref{firstVariationDot}) and (\ref{firstVariationDot2}) yield the \textit{equivariance} of the weak form of the equations:
\begin{equation}
\big<\delta\Pi(\mathcal{T}_Q\boldsymbol\Phi),\textbf{Y}\big>=\big<\mathcal{T}_Q\delta\Pi(\boldsymbol\Phi),\textbf{Y}\big>\quad\textrm{for all $\textbf{Q}\in O(3)$ and admissible $\textbf{Y}$}.
\end{equation}

Now given a mesh with icosahedral symmetry, denoted $S^2_d\subset\mathbb{R}^3$, that is $\textbf{Q}(S^2_d)=S^2_d$ for all $\textbf{Q}\in\mathbb{I}\oplus\mathbb{Z}^c_2$, then the group action (\ref{orthtransform}) is now restricted to $\mathbb{I}\oplus\mathbb{Z}_2^c\subset O(3)$. In view of (\ref{approx1}) and (\ref{approx2}), this subgroup action \quotes{sends} the values of $\rho(\textbf{X}_i)$ and $\phi(\textbf{X}_i)$ at mesh-point location $\textbf{X}_i\in S^2_d$ to $\rho(\textbf{Q}^T\textbf{X}_i)$ and $\phi(\textbf{Q}^T\textbf{X}_i)$, respectively for all $\textbf{Q}\in\mathcal{I}\oplus\mathcal{Z}^c_2$. The icosahedral symmetry of the mesh insures that $\textbf{Q}^T\textbf{X}_i\in S^2_d$. For each $\textbf{Q}\in\mathcal{I}\oplus\mathcal{Z}^c_2$, this indicates a matrix action $\mathcal{T}_Q\textbf{u}$, where $\textbf{u}\in\mathcal{R}^N$ is the vector of unknowns (\ref{unknownFull}) and $\mathcal{T}_Q$ is an $N\times N$ orthogonal matrix. Hence, (\ref{energy_invariance}) is the inheritance of (\ref{PiTF}) on the discretized energy $\hat{\Pi}(\textbf{u},\lambda)$.

\section{Method to compute the $G$-reduced system}
In order to explicitly realize the inherent reduction, we need to express the reduced problem $\textbf{F}_G(\textbf{u},\lambda)$ relative to a basis of the fixed point space $\mathcal{V}_G$. let's define an orthonormal basis for $\mathcal{V}_G$ as
\begin{equation}\label{basis}
\textrm{span}\{\psi_1,\psi_2,\dots,\psi_{mG}\}=\mathcal{V}_G,\quad\textrm{with $\psi_i^t\psi_j=\delta_{ij}$},
\end{equation}
where $mG=dim\mathcal{V}_G$ and $\psi_i$ are called \textit{symmetry modes}.

In Appendix A, we computed $\mathcal{P}$ as a symmetric orthogonal projection from $\mathbb{R}^N$ onto $\mathcal{V}_G$, thus, the basis (\ref{basis}) always exists. Since $\mathcal{V}_G$ is the invariant subspace under $\mathcal{T}$, we have the following identity
\begin{equation}
\mathcal{P}\psi=\psi,\quad\textrm{for all $\psi\in\mathcal{V}_G$}.
\end{equation}

Thus, the basis (\ref{basis}) can be obtained by finding the $dim\mathcal{V}_G$-nontrivial solutions of the following homogeneous system
\begin{equation}\label{eigen_equation}
[\mathcal{P}-\textbf{I}]\psi=\textbf{0},
\end{equation}
These are eigenvector-equations for unit eigenvalues. Projection operator $\mathcal{P}$ has $dim\mathcal{V}_G$ unit eigenvalues. Thus, $\psi_i$ are the $mG$ corresponding eigenvectors. Let $\Psi_G$ be the $n\times mG$ matrix with columns equal to the basis vectors, viz.,
\begin{equation}\label{ProjPsi}
\Psi_G\equiv\big[\psi_1,\psi_2,\dots,\psi_{mG}\big].
\end{equation}

For any vector in the fixed point space $\textbf{u}\in\mathcal{V}_G$,
$\bar{\textbf{u}}\equiv\Psi_G^t\textbf{u}\in\mathbb{R}^{mG}$ is the vector containing the components of $\textbf{u}$ relative to the basis in (\ref{basis}). Thus, the reduced problem relative to the orthonormal basis of $\mathcal{V}_G$ is given by
\begin{equation}\label{reducedF}
\bar{\textbf{F}}_G(\bar{\textbf{u}},\lambda)=\Psi_G^t\textbf{F}(\Psi_G\bar{\textbf{u}},\lambda)=\textbf{0},
\end{equation}
where $\bar{\textbf{F}}_G:\mathbb{R}^{mG}\times\mathbb{R}\to\mathbb{R}^{mG}$. According to Theorem \ref{solutionEquiv}, if $(\bar{\textbf{u}}_0,\lambda_0)$ is a solution point of (\ref{reducedF}), then $(\textbf{u}_0,\lambda_0)\equiv(\Psi_G\bar{\textbf{u}}_0,\lambda_0)$ is a solution point of (\ref{f}). Moreover, there is a unique local branch of solutions of (\ref{reducedF}) through $(\textbf{u}_0,\lambda_0)$.

\section*{Acknowledgement}
This work was supported in part of the National Science Foundation through grant DMS-1312377, which is gratefully acknowledged. We also thank Jim Jenkins,  Chris Earls and Sanjay Dharmaravam for valuable discussions concerning this work.

\bibliography{mybibfile}

\end{document}